\documentclass[fleqn,10pt]{wlscirep}
\usepackage[utf8]{inputenc}
\usepackage[T1]{fontenc}
\usepackage{subcaption}
\usepackage{float}

\usepackage[table]{xcolor}
\usepackage{multicol}
\usepackage{multirow}
\usepackage{pdflscape}
\usepackage{booktabs}
\usepackage{amsmath}
\usepackage{longtable}

\title{Automated versus Human Engagement: Mapping Cognitive Bias Triggers in Online Discourse}

\author[1,*]{Lynnette Hui Xian Ng}
\author[2]{Wenqi Zhou}
\author[1]{Kathleen M. Carley}
\affil[1]{Societal and Software Systems
Carnegie Mellon University, Pittsburgh, PA 15213}
\affil[2]{Information Systems and Technology, Duquesne University, 600 Forbes Ave, Pittsburgh, PA 15282}

\affil[*]{lynnetteng@cmu.edu}


\begin{abstract}
In the digital environment, human attention is frequently guided by cognitive heuristics rather than deliberate evaluation. Since low-credibility narratives often lack substantive factual evidence, their diffusion disproportionally relies on activating these mental shortcut to simulate credibility and capture attention. This study presents a computational framework designed to detect computational triggers through observable data proxies for eight distinct cognitive biases across 3.5 million posts of contested COVID-19 narratives. We demonstrate that automated accounts (bots) embed these triggers more frequently than human users, yielding distinctly source-dependent associations with audience interaction. In bot-authored posts, affective and cognitive dissonance (stance-shifting) triggers are strongly associated with higher engagement, while the deployment of authority and availability (repetition) cues correlates with reduced audience interaction. Furthermore, we identify limits to heuristic compounding: positive engagement correlations with bot-authored content declines when multiple biases are stacked within a single post, whereas human-authored communication remains structurally resilient to high trigger density. By operationalizing psychological heuristics into scalable, measurable data, this work bridges computational social science and cognitive psychology to reveal how source identity (bot/human) shapes the mechanics of information diffusion in digital networks.
\end{abstract}
\begin{document}

\flushbottom
\maketitle
%
%

\section{Introduction}
Social media platforms have transformed how people engage with information, offering instant access to news, opinions, and personal narratives. With billions of posts circulating daily, users must rapidly decide what to read and engage with. These decisions are rarely the product of careful deliberation, but are often guided by judgment heuristics, or mental shortcuts, in order to navigate an overwhelming, fast-paced digital information landscape \cite{metzger2013credibility}. In this attention economy, user focus is a scarce and valuable resource that drives content creators into fierce competition for engagement \cite{menczer2020attention}. As a result, embedding cognitive bias triggers into social media content can intentionally activate these judgment heuristics to influence user perception, memory, and behavior \cite{petty2008persuasion}.

Cognitive bias triggers are utilized extensively in marketing and political messaging. For example, emotionally charged language is frequently used to promote anti-vaccination campaigns \cite{wawrzuta2021characteristics}, ideological and religious issues are amplified through repetitive messaging \cite{danaditya2022curious}, and online advertisements often leverage expert endorsements to lend credibility \cite{xu2022characterizing}.
Such strategies are effective precisely because they activate biases related to authority, availability, and affect, respectively, thereby bypassing deliberative reasoning. Recent frameworks on the psychological factors behind viral online content support this mechanism, showing that content is more likely to spread when it resonates emotionally, signals social identity, or offers morally salient cues \cite{rathje2025psychology}.

In recent years, automated accounts, also known as bots, have played a disproportionately active role in shaping social media discourse \cite{arceneaux2026social}. They are engineered to disseminate information, engage target audiences, express opinions, and ultimately persuade users to take specific actions \cite{chen2022social,ndlela2020social, himelein2021bots,broniatowski2018weaponized}.  Studies estimate that between 5 to 47\% of online users are likely bots, depending on the context and popularity of a given event \cite{tan2023botpercent,fukuda2022estimating}. Furthermore, humans struggle to differentiate between bots and human users approximately 71\% of the time \cite{kenny2024duped}, suggesting that these automated agents have a high degree of believability and persuasive capacity. Bots leverage automation to rapidly produce and amplify content \cite{shao2018spread,dongqi2026expanding}, often embedding themselves in the early stage of narrative building to strategically influence the agenda \cite{li2024social}. This highlights a structural mechanism in information diffusion. Unverified narratives gain traction not merely through algorithmic sorting, but because automated architectures systematically activate the specific cognitive biases, i.e., the natural mental shortcuts, that direct human attention \cite{acerbi2019cognitive}.

While both bots and humans can embed triggers of cognitive biases in their contents, such as emotional language or repetitive framing, to activate cognitive shortcuts, the usage patterns by both groups likely differ. This reliance on heuristic processing is especially critical for the diffusion of unverified or low-credibility narratives. Because such content lacks substantive factual evidence, it is heavily dependent on activating cognitive biases to establish perceived validity and achieve virality \cite{pennycook2019fighting,vosoughi2018spread}. Therefore, studying contested discourse, such as unverified claims surrounding the COVID-19 vaccine, provides an ideal environment to observe exactly how these triggers are deployed. However, existing work has not distinguished between how bots and humans differentially deploy these persuasive tactics, nor compared their relative effectiveness. This paper addresses this gap by comparing bot-authored versus human-authored content through the lens of cognitive bias activation. In doing so, we offer a novel framework for understanding not only what drives online engagement, but why it works, contextualized by the nature of the author of the content.

To this end, we developed a theory-driven computational framework designed to detect and quantify the triggers for eight cognitive biases. We applied this framework to a large-scale corpus of contested narratives about COVID-19 vaccines on X (July 2020 - July 2021) to map how these triggers are distributed, how they co-occur, and how they correlated with user engagement. The analysis reveals distinct behavioral patterns: automated accounts embed these computational triggers significantly more often than human users, which corresponds with divergent patterns of audience interaction. By operationalizing psychological heuristics into measurable data, this study bridges cognitive psychology and computational social science to clarify the mechanics of automated information dissemination.

\section{From Cognitive Heuristics to Computational Triggers}
\label{sec:heuristics}
To quickly process the overwhelming volume of content on social media, people operate under bounded rationality, relying on mental shortcuts known as heuristics \cite{tversky1974judgment}. Psychological research categorizes these heuristics into three main types: Representativeness, Availability, and Anchoring \cite{tversky1974judgment}. While these heuristics generally facilitate the efficient processing of information, the high-speed environment of social media can cause them to systematically misfire, resulting in cognitive biases. Online narratives frequently tap into this vulnerability. By embedding specific cues, which we term \textit{bias triggers}, content creators can activate these heuristics to subtly steer audiences toward preferred interpretations  \cite{braca2023developing}. Detecting these triggers is essential, as they heavily influence engagement with low-credibility or controversial information \cite{sun2024heuristic}. Automated accounts are uniquely positioned to provide this structural support at scale, acting as the primary delivery mechanism for heuristic-driven content. 

Drawing on this psychological framework, we introduce the concept of \textit{computational triggers} to detect these cues at scale. We analyzed the conceptual definitions of eight cognitive biases as they manifest on social media and identified their measurable data proxies for their triggers. These computational features include linguistic patterns, author identity, and temporal dynamics. We provide detailed discussions and explanations of how we operationalize these triggers specifically for our target platform, X, in the subsequent Methodology section (\autoref{sec:operation_bias}). 

Judgment by Representativeness assesses the likelihood of an outcome based on how closely it resembles a known category or stereotype. Within this heuristic, we examine two biases. \textit{Homophily Bias} refers to the tendency to favor information from similar individuals. Online, Homophily Bias is triggered when users preferentially interact with content created by accounts sharing their own identities. \textit{Authority Bias} highlights the inclination to perceive messages from reputable sources as credible. On social media, Authority Bias is triggered and computationally identified when established authority figures are tagged or mentioned within a post. 


Judgment by Availability occurs when people estimate an event’s likelihood based on how easily examples come to mind. \textit{Availability Bias} drives users to adopt beliefs based on perceived popularity, triggered by highly visible engagement metrics, such as retweet volume, that signal widespread endorsement \cite{boyd2010tweet}. The \textit{Illusory Truth Effect} refers to the tendency to believe repeated information regardless of factual accuracy. This can be triggered when content creators strategically repeat identical messages to shape perception \cite{danaditya2022curious}. \textit{Affect Bias} and \textit{Negativity Bias} triggers use broad and negative emotions, respectively, to heighten the perceived validity of a claim \cite{tversky1974judgment}. Because emotional resonance accelerates the spread of controversial information on social media \cite{horner2021emotions}, we quantify these emotional triggers using psycholinguistic analysis \cite{tausczik2010psychological}. 

Finally, judgment by Anchoring heuristic evaluates new information by referencing prior anchors, such as personal experiences or existing beliefs. We measure two biases in this category. \textit{Confirmation Bias} favors information that strengthens preexisting beliefs, manifested online as selective exposure \cite{zhao2020promoting}. We detect its computational trigger when online accounts continually post contents that maintains a singular stance.
\textit{Cognitive Dissonance} anchors individuals to their communities' prevailing views. To reduce mental discomfort, people often shift their opinions to conform, leading to echo chambers \cite{mocanu2015collective}. Borrowing from past work \cite{ng2022pro}, we measure the trigger for this dissonance-drive conformity by tracking shifts in a user's expressed stance that bring them into alignment with their networks.

\section{Methodology}
\subsection{Dataset Construction} We acquired the COVID-19 data from a published dataset \cite{ng2025global}. We selected this dataset because it contains contested and conflicting narratives at scale during this public health crisis. Such narratives are particularly suitable for detecting cognitive triggers and studying their engagement with online users \cite{wu2025people}, because the ambiguity and emotional salience of such content naturally elicits heuristic-driven processing by the human psychology \cite{pennycook2021psychology}. This dataset collected tweets that were publicly posted by users on X regarding the COVID-19 pandemic. No attempt was made to retrieve any user's private tweets. These tweets were obtained from using the X's official API.

To focus our analysis on contested narratives, we filtered the corpus for tweets containing contested narratives, such as ``salt solution can cure covid19", ``this is too perfect a bioweapon to have occurred naturally". The extraction of such narratives was performed using cosine similarity comparisons of TwHIN-BERT-based vector embeddings and matching tweet content against a vector space of expertly annotated unverified or low-credibility COVID-19 narratives from a well-adopted annotated dataset of contested COVID-19 X posts \cite{memon2020characterizing}. From an original dataset of 4,179,124,820 tweets, this produces a subset of 3,550,960 tweets with contested narratives (0.08\%) for analysis.

To categorize author accounts, we applied the tier-based BotHunter algorithm, which calculates an automation probability score between 0 and 1 \cite{beskow2018bot}. Based on a large-scale statistical analysis of this algorithm \cite{ng2022stabilizing}, accounts scoring above or equal to 0.7 were classified as automated accounts (bots), with the remainder classified as human users. 
Bot accounts represented only 3.1\% of the total users in this original corpus, but they produced 41.2\% of the total tweets.

Details regarding data curation, contested tweets extraction, and parameters and performance metrics (including accuracy and precision validation) for the BotHunter algorithm are provided in the Supplementary Materials \autoref{supp:dataset_construction}. 

\subsection{Operationalizing Bias Triggers on X} 
\label{sec:operation_bias}
To systematically identify the presence of cognitive bias triggers on the X platform, we developed a rule-based computational framework grounded in psychological theories outlined previously. We operationalized triggers for eight specific biases: Availability, Confirmation, Cognitive Dissonance, Affect, Negativity, Authority, Homophily, and the Illusory Truth Effect. These eight specific biases were selected based on a preliminary literature survey of frequently exploited heuristics during crisis events. The empirical literature references and the cognitive biases studied are presented in Supplementary Material \autoref{tab:bias_literature}. Crucially, these biases manifest via behaviors that can be rigorously operationalized into measurable computational proxies. \cite{boyd2010tweet,stella2018bots,lee2024correcting}.  

The computational rules were derived from a manually annotation of a random n=800 sample of bot-authored tweets. Following standard procedures, two domain experts independently annotated the sample, with a third resolving disagreement.  Annotators identified lexical (e.g., reference authority sources, identity matching), syntactic structural (e.g., tweets are at least 80\% similar to each other) and quantitative patterns (e.g., number of retweets, number of emotion words) serving as bias triggers. Subsequently, we validated the computational framework by comparing its automated identification of bias triggers against the human annotation consensus. Further details on the human annotation process, including inter-annotator reliability, and algorithm validation are reported in \autoref{sec:human_annotation}.

To ensure that we captured systematically embedded triggers rather than baseline conversational noises, we established a dynamic threshold, $k$, defined as the $\text{ceiling}(\mu+\sigma)$ of the cue distribution across all the tweets, a threshold technique referenced from past work\cite{scalco2026detect,mredula2022review}. The descriptive statistics of cognitive cues along with engagement metrics are included in Supplementary Materials \autoref{tab:desc_stats}. Our framework detects these triggers across three distinct dimensions:

\begin{enumerate}
    \item \textbf{Tweet-level Triggers: } Homophily and Authority Bias triggers are rooted in Social Identity Theory, which outlines that interpersonal relations are closely related to the identity groups that a person affiliates with \cite{tajfel2004social}. A Homophily Bias trigger is flagged when an author shares tweets from a user with a matching group affiliation. An Authority Bias trigger is flagged when an author claims authority or explicitly references/tags an established authority figure, either through text or @mention tags. While psychological Authority Bias is highly complex, we utilize explicit tagging of established figures as our computational proxy because it represents the most direct, observable mechanism an author can use to visually borrow credibility on the X platform. Affiliations and authority statuses were identified against a 2023 U.S. population survey, which provided occupational and social groups (``student", ``teacher", ``dad") and authority statuses (``president", ``CEO") \cite{smith2016mean}. Computationally, each X user is represented through these affiliation and authority status through parsing their metadata (bio, description) to identify the words provided in the population survey.

    \item \textbf{User-level Triggers: } Availability Bias triggers simulate widespread popularity and are flagged when a user shares (i.e., retweets, quotes) the exact same tweet $k\geq3$ times. The Illusory Truth Effect trigger, which capitalizes on repetition, is flagged when a single user posts $k\geq3$ tweets that posses a semantic similarity of $r\geq80\%$ . Semantic similarity was measured through cosine similarity on BERT sentence vector embeddings, and the threshold $r$ was adopted from previous studies about similarities of  social media posts \cite{ng2022cross,novo2023explaining}.

    \item \textbf{Temporal and Network-Level Triggers}: Cognitive Dissonance triggers are via network conformity. We constructed an interaction network graph $G=(V,E)$ for each user (the ego). The vertices $V$ consists of the ego and all other users that interacted with the ego through retweet, @mentions, quotes, and replies. Graph edges $E$ between two nodes $v_i$ and $v_j$ represents that the $v_i$ interacted with $v_j$. A trigger is flagged if the ego's initial stance diverges from their network's majority stance, but the stance of their subsequent message shifted to match the network's consensus. Confirmation bias triggers are identified sequentially, flagged when an author posts $k\geq3$ consecutive tweets that maintained an identical stance on COVID-19 vaccines, indicating a structural reinforcement of a singular narrative.
\end{enumerate}

To ensure computational robustness, each heuristic was iteratively tested and refined through manual inspection of a random 20-tweet validation set. The final detection schema is detailed in \autoref{tab:bots_bias_results}.

\newgeometry{margin=1cm} 
\begin{landscape}

\begin{table}[h]
    \centering
    \begin{tabular}{|p{4cm}|p{7cm}|p{7cm}|p{2.5cm}|}
    \hline
        \textbf{Cognitive Biases}& \textbf{Example Observation in Dataset}& \textbf{Computational Definition of Triggers}& \textbf{Scope} \\ \hline        
        \multicolumn{4}{|c|}{\textbf{Representativeness Heuristics}} \\ \hline
        Homophily Bias & Author @EasyWorldNews writes: \emph{``@globalfirstnews Readers' poll: if you are offered a Covid vaccination, will you accept? "} & User shares content from a user with a matching demographic or group affiliation.& Tweet \\ \hline
        Authority Bias & Author retweets: \emph{``RT @DrEricDing: Dangerous anti-vaccine \& far-right groups shut down Dodger Stadium's mass \#COVID19 vaccination site [...]"} & User explicitly tags/references established authority figures, or possesses verified authority status.& Tweet \\ \hline 
        \multicolumn{4}{|c|}{\textbf{Availability Heuristics}} \\ \hline
        Availability Bias & Author retweets three times: \emph{``RT @User1: India fastest country to cross 1 million Covid-19 vaccinations, 25 lakh doses administered so far: Government"} & User shares (retweets or quotes) the exact same tweet $k\geq3$ times.& User \\ \hline 
        Illusory Truth Effect & Author shares the tweet \emph{``The fastest way to end the \#COVID19 pandemic is to make safe an  d effective \#vaccines available to everyone on the planet!"} three times, each tagging different users. & User posts $k\geq3$ distinct tweets that share an $r\geq80$\% semantic similarity.& User \\ \hline 
        Affect Bias & \emph{``Apartheid Israel is withholding the coronavirus vaccine from Palestinians, whilst simultaneously bombing their hospital"}  & Tweet contains $k\geq3$ emotional words based on psycholinguistic markers.& Tweet\\ \hline 
        Negativity Bias & \emph{``This is a disaster, and it's getting worse!!: Inside Pfizer's feverish rush to bring a Covid-19 vaccine to market in record time"} & Tweet contains $k\geq2$ negative words based on psycholinguistic markers.& Tweet \\ \hline 
        \multicolumn{4}{|c|}{\textbf{Anchoring Heuristics}} \\ \hline
        Cognitive Dissonance & First tweet shows anti-vaccine sentiments: \emph{``[...]I will never have the vaccine ever, [...]\#NoVaccine" to ``March 2020 wwas the hardest. March 2021 slightly better due to \#vaccines"}; second tweet shows pro-vaccine sentiment: \emph{``Protect your community. \#getthevaccine"} & User's initial stance regarding Covid-19 vaccine diverges from their interaction network's majority, but a subsequent tweet shifts to align with the network consensus.& Temporal, Network\\ \hline
        Confirmation Bias & Three tweets from the same user shows anti-vaccine sentiment: \emph{``Immediately add immunity-building/antiinflammatory/anti-viral garlic/Vitamin D3 to the treatment mix!}; \emph{Garlic cuts colds by 50\% (COVID-19 is a form of a cold)"}; \emph{``per the Israelis, 2000-5000 IUs of daily D3 cuts COVID+ cases 50\% also."} & User posts $k\geq3$ consecutive, non-identical,  tweets maintaining an identical stance towards the COVID-19 vaccine; or the tweet contains at least 2 distinct sentences of an identical stance towards the vaccine.& Temporal\\ \hline
    \end{tabular}
    \caption{\textbf{Computational Framework for Bias Triggers. }}
    \label{tab:bots_bias_results}
\end{table}
\end{landscape}
\restoregeometry

\subsection{Associating Triggers with Tweet Engagement}
To evaluate the relationship between these embedded triggers and subsequent content amplification, we conducted an Ordinary Least Squares (OLS) regression. We posit that computational triggers in both bot-authored and human-authored tweets activate cognitive biases in receivers, which theoretically increases the likelihood of active engagement and narrative propagation \cite{wu2025people}. Following established literature, we utilized standard platform engagement metrics as quantitative proxies for content engagement and persuasiveness \cite{ibrahim2015persuasive,kim2018rumor}, including the number of favorites, the number of retweets, the number of replies, and the number of quotes.

For each metric (favorites, retweets, replies, and quotes), we modeled the log-transformed engagement count as a function of the presence of the bias triggers. The log transformation was applied to correct for the skewed distribution typical of social media engagement and improve interpretability. Eight binary dummy variables were included for each tweet, indicating the presence or absence of each trigger. To isolate group-specific behavioral differences, regressions were run individually for bot-authored and human-authored subsets. Therefore, one equation was constructed for bots, and another one for humans. Finally, we correlated the total number of bias triggers within a single tweet with the log-transformed engagement metrics to test the effect of the quantity of triggers. 



\section{Results}
We characterized a final filtered corpus of 3,550,960 tweets containing unverified COVID-19 narratives to assess the presence of computational bias triggers. Within this filtered dataset, 65.47\% of all human-authored tweets contained unverified or low-credibility narratives, compared to a near-universal 98.60\% among bot-authored tweets. 

For each tweet, we computationally flagged the presence of bias triggers using our rule-based proxies for the representativeness, availability, and anchoring heuristics. Using this labeled dataset, we analyzed how the distribution and co-occurrence of these triggers differed between human and bot accounts, and how their presence correlated with subsequent audience engagement.

\subsection{Differential Usage of Bias Triggers by Source}
\begin{figure}[h]
\centering
\includegraphics[scale=0.3]{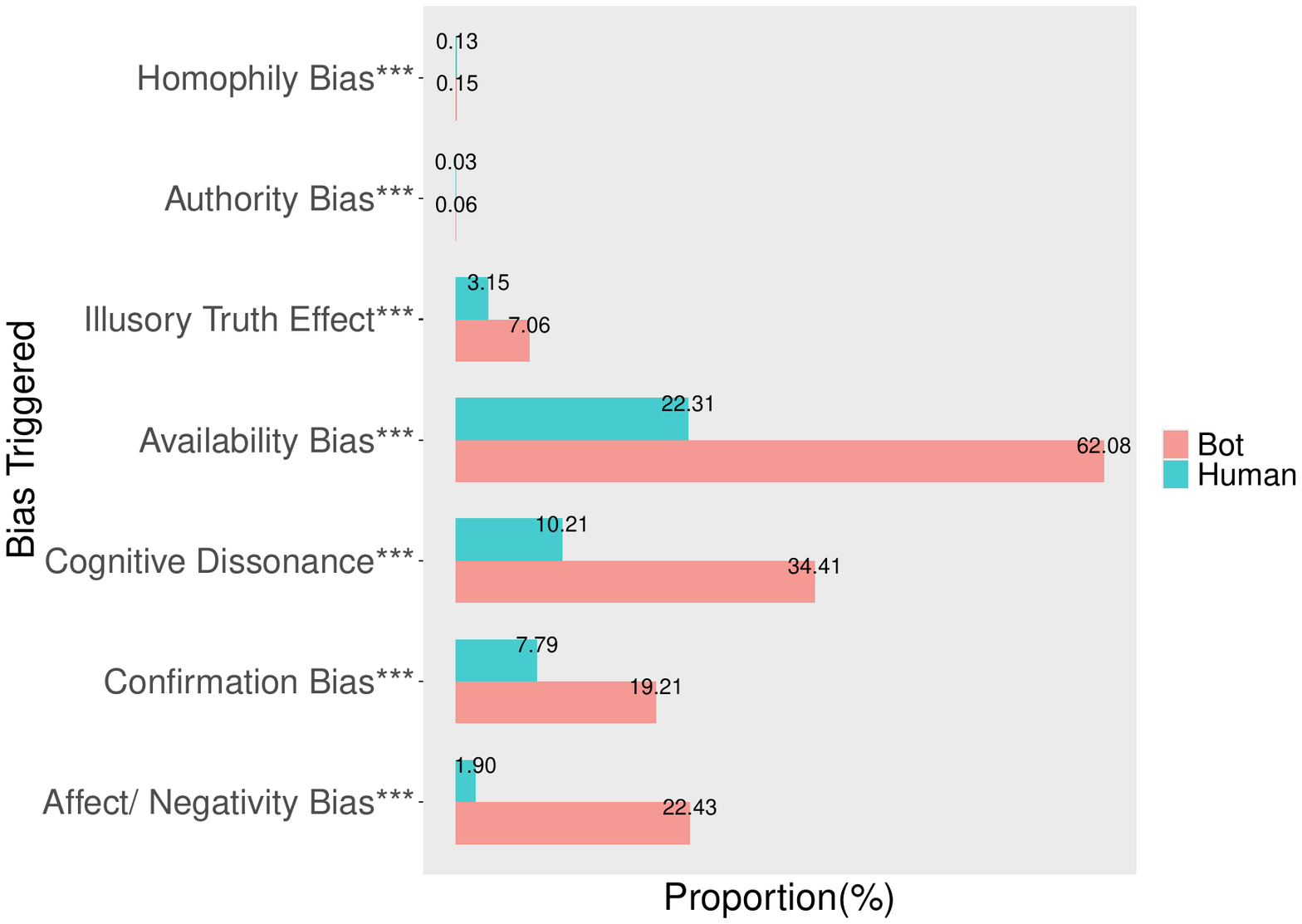}
\caption{\textbf{Distribution of the Bias Triggers.} This illustrates the percentage of tweets containing computational triggers for cognitive biases, categorized by  user type. Affect and Negativity Bias were merged into a single construct due to the complete empirical overlap.}
\label{fig:biasproportion}
\end{figure}

We first compared the proportion of tweets that contained bias triggers across user types. As illustrated in  \autoref{fig:biasproportion}, bots embedded bias triggers in their tweets at a rate 4.71$\pm$4.42 times higher than humans. The majority of human tweets (54.16\%) did not contain any measurable bias triggers. In contrast, 80.19\% of bot tweets contained at least one, indicating that automated architectures systematically leverage these heuristic cues to drive early-stage engagement \cite{safadi2024effect}. A proportion $z$-test confirmed that bot tweets contained significantly more bias triggers than human tweets across all eight categories. The narrowest margin appeared in Authority Bias; however, while the relative percentages were similar (0.06\% bots, 0.03\% humans), the ratio of the absolute volume of bot tweets to human tweets utilizing this trigger is 1.4:1.0. That is, there were 52,666 more bot tweets than human tweets that utilized this trigger.

Second, the two most frequently triggered biases among both groups were Availability Bias (62.08\% for bots, 34.41\% for humans) and Cognitive Dissonance (22.31\% bots, 10.21\% humans). The prevalence of these triggers in bot content highlights two computational mechanisms used to simulate group cohesion. First, bots exhibit high rates of stance-shifting (Cognitive Dissonance triggers) to align with dominant network opinions, a manufactured consensus that algorithms often reward\cite{ng2022pro}. Second, bots heavily rely on high-frequency repetition of identical messages (Availability Bias triggers) to artificially inflate the perceived popularity of a narrative. Together, these structural strategies reinforce the illusion of truth and forge a seemingly united collective belief \cite{prollochs2023mechanisms}.

Third, triggers relying on emotional arousal (Affect and Negativity Bias) were rare in human tweets (1.90\%). However, they appeared in 22.43\% of bot tweets.  Emotional content is a powerful amplifier of online information \cite{chen2021persuasion}. Negative emotional framing increases message memorability by bypassing analytical processing, a heuristic mechanism that automated broadcasting systematically targets. Interestingly, the occurrence rate for Affect Bias and Negativity Bias were mathematically identical across both groups. This indicates that when low-credibility COVID-19  tweets employed affective language, they relied exclusively on negative emotional cues. This aligns with prior research demonstrating that viral news tends to be negatively valenced \cite{al2019viral} and disproportionately disseminated by bots \cite{stella2018bots}. Given this complete empirical overlap, we merged these two categories into a single construct,  ``Affect/Negativity Bias'', for all subsequent analyses.

\subsection{Distinct Co-Occurrence Patterns of Bias Triggers}
\begin{figure*}[tbhp]
\centering
\includegraphics[scale=0.5]{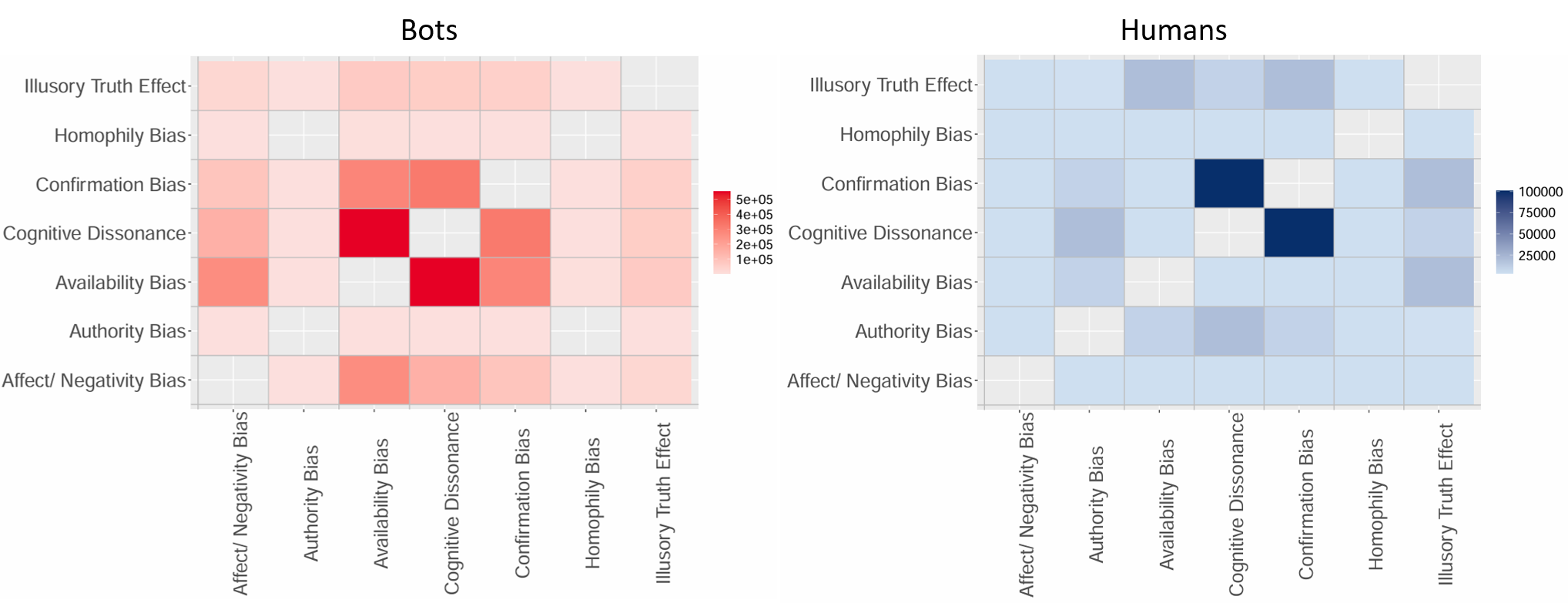}
\caption{\textbf{Co-Occurrence of Bias Triggers.} The heatmap is color-coded according to the prevalence of concurrent bias triggers within individual tweets.}
\label{fig:cooccurrence}
\end{figure*}

Beyond analyzing the presence of bias triggers in their singularity, we also observe that a tweet, or set of tweets, can contain multiple bias triggers.
\autoref{fig:cooccurrence} illustrates the co-occurrence of bias triggers within individual tweets. Darker shades indicate frequent concurrence, implying that specific combinations of triggers are selectively embedded to compound their persuasive impact. This phenomenon is supported by dual-process psychological theories that persuasion is more effective when multiple cognitive shortcuts are activated simultaneously \cite{petty2008persuasion}.

Bot and human tweets exhibit distinctly different paring strategies. Bot content displays highly frequent and diverse trigger combinations. Notably, the most common bot pairing, Cognitive Dissonance and Availability Bias, was rarely observed in human tweets. This echoes the observation from past work that bots frequently shift their stance to align with dominant network opinions and then amplify that new stance via repeated quoting or retweeting, a behavior that contrasts with typical human tendencies \cite{ng2022pro}. Conversely, the most frequent human pairing was Confirmation Bias and Cognitive Dissonance. This suggests that when humans adjust their views to match their social environment, they reinforce their revised stance through original, continuous posting rather than merely sharing existing content. Both strategies seek to reinforce the revised position and construct a coherent narrative trajectory, albeit with different strategies: bots via algorithmic amplification, humans through organic expression.

Another frequent bot pairing, Affect/Negativity Bias and Availability Bias, reflects the automated amplification of emotional content. This aligns with rumor propagation theory, which posits that narratives combining emotional charge with high familiarity are more readily accepted \cite{knapp1944psychology}. However, humans exhibit minimal co-occurrence between Affect/Negativity triggers and other biases', suggesting that human emotional expression tends to emerge organically rather than as part of a compound persuasive strategy \cite{pennebaker2003psychological}. Finally, some pairings like Authority Bias and Homophily Bias appeared sporadically in human tweets but were virtually absent in bot contents. This pattern aligns with social signaling theory: like-minded people occasionally invoke shared authorities to reinforce genuine group identity, whereas bots tend to avoid specific in-group appeals. Instead, they opt for broad and generic messaging strategies to engage with diverse audiences \cite{marwick2011tweet}.

\subsection{Source-Dependent Engagement Effects}
\begin{figure*}[tbhp]
\centering
\includegraphics[scale=0.5]{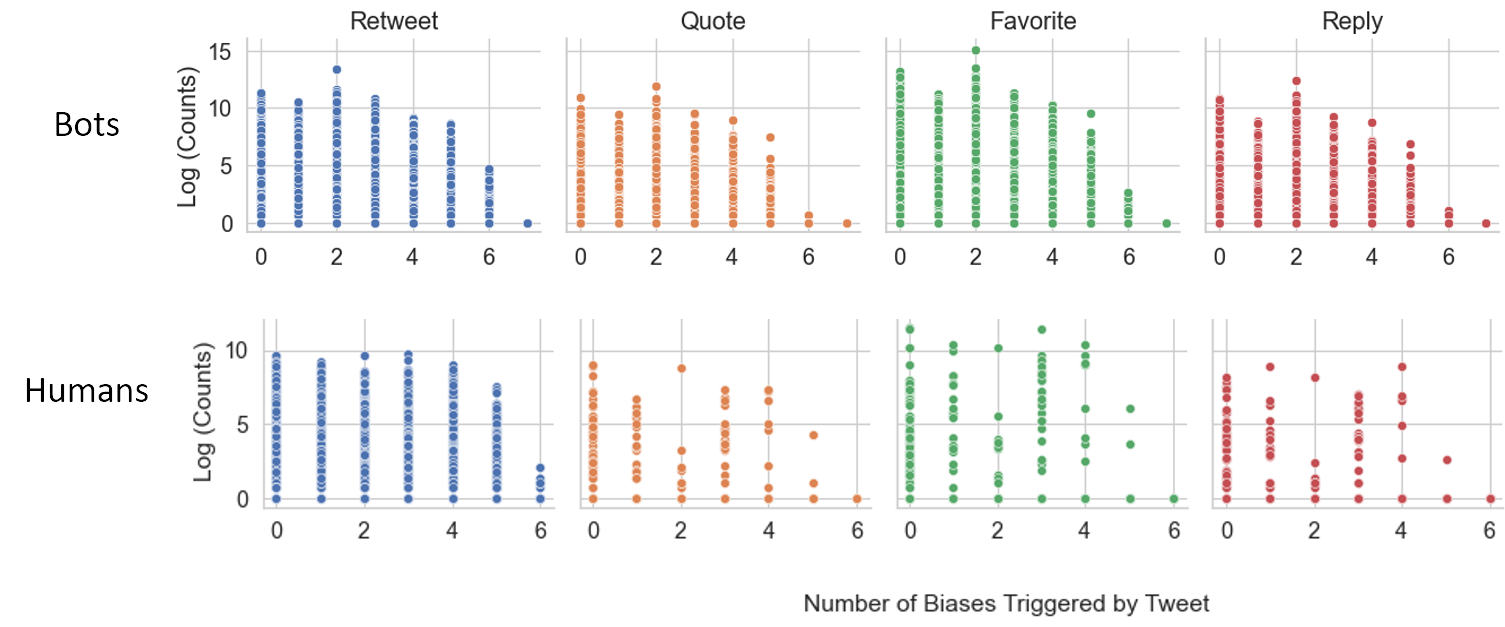}
\caption{\textbf{Associations between the Number of Bias Triggers and Tweet Engagement}}
\label{fig:num_bias_engagement_corr}
\end{figure*}

We next examined how the volume and type of bias triggers correlated with subsequent audience engagement. \autoref{fig:num_bias_engagement_corr} visualizes the distribution of log-transformed engagement metrics relative to the number of bias triggers per tweet. Bot tweets exhibit a clear and relatively smooth trend: engagement rises with the addition of triggers, peaking at approximately two triggers, before gradually declining. This suggests a potential ``sweet spot'' for automated cognitive activation for bot operation, where a moderate number of bias triggers optimally stimulate engagement. Beyond this threshold, the decline likely reflects cognitive overload or a perception of spam, where excessive computational cues cause audiences to disengage  \cite{muhammed2022disaster,fu2020social}. Human tweets, however, maintained stable engagement with increasing number of triggers until the tweet contained more than four triggers. Since humans sometimes have trouble distinguishing whether a tweet was authored by a bot or a human \cite{kenny2024duped}, further research is required to determine why this happens. There could be other cues that humans are putting into the text, like contextual cues or interpersonal conversational cues \cite{zade2024reply,ng2025global}, that caused this effect.

To isolate the impact of specific triggers, we utilized OLS regression models by user type (humans and bots). The results are illustrated in \autoref{fig:mainfigure}. The full report of the regression results can be found in Supplementary Materials \autoref{tab:regression}. The regression revealed several divergence between human and automated content.

First, the presence of bias triggers in bot tweets consistently demonstrated statistically significant associations with engagement, particularly for low-effort, ``shallow'' interactions like favorites and retweets. The triggers showed minimal influence on replies or quotes, which required higher cognitive deliberation. In contrast, human tweets demonstrated practically negligible or economically insignificant effects from these same triggers. This divergence indicates that shallow engagement mechanisms are highly susceptible to the heuristic-driven responses that bots systematically exploit \cite{stella2018bots, shao2018spread}. Human content, reliant on richer semantic meaning, requires more substantive cognitive alignment from readers, making isolated bias triggers insufficient predictors of engagement \cite{marwick2011tweet}.

When embedded in bot tweets, three cognitive triggers, Affect/Negativity Bias, Cognitive Dissonance, and Confirmation Bias, consistently correlated with increased engagement across all metrics. Affect/Negativity triggers correlated with up to 4\% higher engagement, thriving in an attention economy built for emotional resonance \cite{rathje2025psychology}. Cognitive Dissonance triggers (network stance-shifting) were associated with up to 9\% higher engagement, successfully resolving narrative tension for the audience. Confirmation Bias triggers (stance repetition) corresponded with up to 3\% higher interaction by continuously affirming preexisting beliefs and in-group alignment. Through the lens of dual-process theory, these triggers operate primarily via System 1 processing, prompting fast, emotion-driven interactions regardless of factual accuracy \cite{evans2013dual, chaiken1980heuristic}, rather than System 2 processing which is slow, deliberate analysis \cite{kahneman2011thinking}.

In contrast, triggers for Homophily, Availability, and Authority Biases demonstrated a disengaging effect in bot content. Availability Bias triggers (excessive identical resharing) emerged as the strongest negative predictor, associated with up to a 28\% decrease in favorites and a nearly 33\% decrease in retweets, likely due to audiences penalizing highly visible spam behavior \cite{varol2017online}. Authority Bias triggers in bot tweets reduced favorites by 22\% and retweets by 25\%. This reflects a well-documented cultural distrust of explicit authority figures during the COVID-19 pandemic \cite{trent2022trust}\cite{min2009cultural}, a sentiment bots inadvertently amplified by using generic authority tags \cite{grimme2018changing}.

Interestingly, Authority Bias had the exact opposite effect when deployed by humans, actively increasing retweet engagement. This aligns with communication accommodation theory: readers respond favorably to authority heuristics when they trust the human author contextualizing them \cite{gallois2005communication, steinmetz2021liking}. Humans likely understand their audience and reference appropriately selected authorities other than generic authority figures \cite{steinmetz2021liking}.

In sum, these findings underscore that a trigger's association with user engagement is heavily dependent on its source. In automated accounts, computational triggers are deployed systematically and correlate with high volumes of shallow engagement, whereas human communication relies on a more nuanced, organic integration of heuristic cues.

\begin{figure}[tbhp]
\centering
\includegraphics[scale=0.5]{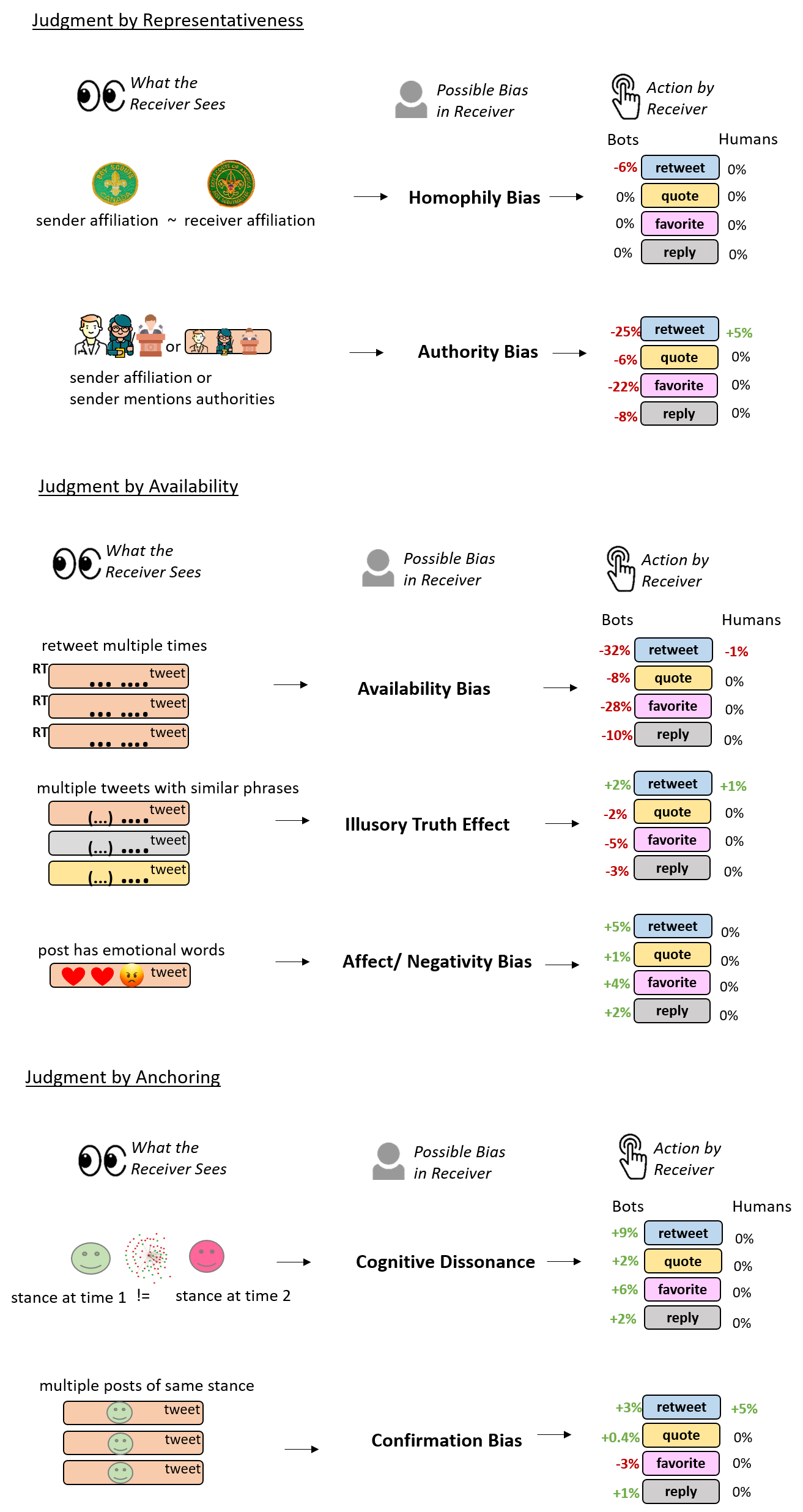}
\caption{Estimated association between bias triggers and tweet engagement. Value represent the percentage change in engagement metrics associated with the presence of each computational trigger. All non-zero estimates are statistically significant ( $p<0.001$); 0\% indicates either statistically insignificant ($p>0.05$) or practically insignificant (value<0.01\%). }
\label{fig:mainfigure}
\end{figure}

\section{Discussion}
Our work investigating the effect of cognitive bias triggers on online engagement bridges computational social science and cognitive psychology. It provides insights into both the structural detection of unverified narratives and the cognitive vulnerabilities that facilitate their spread. By contrasting automated and human behaviors at scale, we demonstrate that automated accounts (bots) systematically deploy computational bias triggers to amplify low-effort engagement, such as favorites and retweets. In contrast, human-driven communication shows weaker associations with these isolated triggers, reflecting the more nuanced, relational dimensions of human persuasion. These findings emphasize that contested narratives thrive not just because of their content, but due to who disseminates them and how they are structurally framed.

This study offers three key contributions to our understanding of online information dissemination. First, we operationalized theories of judgment heuristics for the social media context, developing a computational framework capable of automatically detecting proxies for eight distinct cognitive biases within COVID-19 discourse on X. Crucially, this framework extends beyond extracting isolated lexical cues (e.g., Affect or Negativity Bias) to identify complex structural patterns across multiple posts (e.g., Availability and Confirmation Biases) and dynamic network interactions over time (e.g., Cognitive Dissonance). This approach provides a scalable tool for analyzing how cognitive vulnerabilities are computationally targeted during other high-profile social events.

Second, we reveal systematic differences in how automated and human accounts deploy these triggers. Bots and humans exhibit different profiles regarding the frequency, selection, and combination of bias proxies. Bots embed triggers in a substantially larger proportion of their content, pointing to a structurally embedded strategy for narrative amplification. Most notably, bots disproportionately leverage the Availability Bias trigger, embedding high-frequency repetition in 62\% of their posts compared to only 22\% for humans. Similarly, emotional arousal triggers (Affect/Negativity Bias) appear in roughly one-fifth of automated posts, but in merely 2\% of human posts. These disparities highlight the algorithmic advantages of automation: API access enables the effortless scaling of repetition and the systematic insertion of affective linguistic cues far beyond human physical capacity.
 
Third, our analysis demonstrates a strongly source-dependent relationship between bias triggers and audience engagement. Automated accounts are significantly more likely to compound multiple triggers within a single post, and these combinations correlate strongly with increased engagement. Humans rarely stack triggers, and when they do, the engagement effects are minimal. Furthermore, the effectiveness of specific triggers varies drastically. While bot-embedded Affect/Negativity and Cognitive Dissonance triggers successfully drive interaction, their use of Availability and Authority cues tends to suppress it. This suggests that automated accounts effectively capitalize on rapid, emotionally arousing, and stance-shifting tactics to capture attention. In contrast, human discourse relies more on stable authority and consistency markers—strategies that require deeper cognitive alignment from readers and may inherently limit viral spread. 

In sum, these observations provide insights on the strategic and inorganic nature of automated information operations. The formulaic co-occurrence patterns we observed, such as pairing network stance-shifting (Cognitive Dissonance) with high-volume repetition (Availability Bias), indicate a systemic optimization for heuristic impact rather than genuine discourse \cite{ng2022pro}. While prior studies note that digital ecosystems are vulnerable to affective and familiarity heuristics, our results provide a structured, empirical mapping of exactly how automated agents mechanically execute these strategies, via repetition, stance manipulation, and sentiment targeting, to artificially inflate narrative virality. From an applied perspective, this framework informs structural approaches to mitigating the spread of unverified narratives. Because traditional moderation relies heavily on resource-intensive fact-checking, tracking the behavioral footprint of compounded heuristic proxies, such as the simultaneous stance-shifting and high-frequency repetition characteristic of automated operations, offers a scalable alternative. This approach enables platform architectures to systematically introduce algorithmic friction against inorganic amplification, entirely independent of a narrative's factual veracity.

This study does, however, present certain limitations. First, the analysis is restricted to unverified or low-credibility COVID-19 vaccine narratives and eight specific heuristics; trigger efficacy and distribution likely vary across different event topics. However, we utilize this period as a canonical, illustrative case study of high-stakes crisis informatics. The computational triggers mapped here—relying on fundamental human heuristics—remain highly relevant to contemporary online information ecosystems, including those driven by modern generative AI, as the underlying psychological vulnerabilities they exploit do not change. Second, while our classification framework achieved robust agreement ($\sim$62.7\%) with expert annotations, the presence of a computational trigger remains a proxy. True psychological intent is not directly observable. Nevertheless, this work opens promising avenues for future research. Subsequent studies could integrate network topology to determine how an account's structural position amplifies or mitigates heuristic impact, or pair these computational findings with controlled psychological experiments to validate exactly how human users cognitively process these algorithmic cues.

\section*{Conclusion}
In this article, we examined how cognitive bias triggers embedded in online discourse shape engagement dynamics across automated and human users. By operationalizing eight psychological heuristics into measurable computational proxies, we demonstrate that the engagement of contested narratives is not only a function of content credibility, or lack of, but also the strategies that can potentially trigger human cognitive biases. Our analysis of 3.5 million posts highlights distinct roles of two user types within the online ecosystem: automated accounts function as heuristic-targeting agents engineered for influence at scale, while human communication remains socially embedded and contextually constrained. This work contributes to interdisciplinary efforts in computer science and cognitive psychology by providing a scalable framework for detecting cognitive bias triggers that contribute to viral narrative engagement online.

\bibliography{references}

@article{ng2025global,
  title={A global comparison of social media bot and human characteristics},
  author={Ng, Lynnette Hui Xian and Carley, Kathleen M.},
  journal={Scientific Reports},
  volume={15},
  number={1},
  pages={10973},
  year={2025},
  publisher={Nature Publishing Group UK London}
}

@article{zade2024reply,
  title={To reply or to quote: Comparing conversational framing strategies on Twitter},
  author={Zade, Himanshu and Williams, Spencer and Tran, Theresa T and Smith, Christina and Venkatagiri, Sukrit and Hsieh, Gary and Starbird, Kate},
  journal={ACM Journal on Computing and Sustainable Societies},
  volume={2},
  number={1},
  pages={1--27},
  year={2024},
  publisher={ACM New York, NY}
}

@book{hastie2009elements,
  title={The elements of statistical learning: data mining, inference, and prediction},
  author={Hastie, Trevor and Tibshirani, Robert and Friedman, Jerome H and Friedman, Jerome H},
  volume={2},
  year={2009},
  publisher={Springer}
}

@article{combrisson2015exceeding,
  title={Exceeding chance level by chance: The caveat of theoretical chance levels in brain signal classification and statistical assessment of decoding accuracy},
  author={Combrisson, Etienne and Jerbi, Karim},
  journal={Journal of neuroscience methods},
  volume={250},
  pages={126--136},
  year={2015},
  publisher={Elsevier}
}

@incollection{tajfel2004social,
  title={The social identity theory of intergroup behavior},
  author={Tajfel, Henri and Turner, John C},
  booktitle={Political psychology},
  pages={276--293},
  year={2004},
  publisher={Psychology Press}
}

@book{kahneman2011thinking,
  author    = {Kahneman, Daniel},
  title     = {Thinking, Fast and Slow},
  year      = {2011},
  publisher = {Farrar, Straus and Giroux},
  address   = {New York},
  isbn      = {9780374275631}
}

@inproceedings{novo2023explaining,
  title={Explaining BERT model decisions for near-duplicate news article detection based on named entity recognition},
  author={Novo, Anne Stockem and Gedikli, Fatih},
  booktitle={2023 IEEE 17th International Conference on Semantic Computing (ICSC)},
  pages={278--281},
  year={2023},
  organization={IEEE}
}

@article{jalbert2025perceived,
  title={Who is perceived to be an expert on COVID-19 vaccines on social media? Biomedical credentials confer expertise, even among vaccine-hesitant and conservative observers},
  author={Jalbert, Madeline and Harris, Mallory and Williams, Luke},
  journal={Information, Communication \& Society},
  volume={28},
  number={4},
  pages={669--687},
  year={2025},
  publisher={Taylor \& Francis}
}

@article{li2025tiktok,
  title={TikTok’s political landscape: Examining echo chambers and political expression dynamics},
  author={Li, Yanlin and Cheng, Zicheng and Gil de Z{\'u}{\~n}iga, Homero},
  journal={new media \& society},
  pages={14614448251339755},
  year={2025},
  publisher={SAGE Publications Sage UK: London, England}
}

@article{haselswerdt2024echo,
  title={Echo chambers or doom scrolling? Homophily, intensity, and exposure to elite social media messages},
  author={Haselswerdt, Jake and Fine, Jeffrey A},
  journal={Political Research Quarterly},
  volume={77},
  number={1},
  pages={199--212},
  year={2024},
  publisher={SAGE Publications Sage CA: Los Angeles, CA}
}

@article{opitz2024closer,
  title={A closer look at classification evaluation metrics and a critical reflection of common evaluation practice},
  author={Opitz, Juri},
  journal={transactions of the association for computational linguistics},
  volume={12},
  pages={820--836},
  year={2024},
  publisher={MIT Press 255 Main Street, 9th Floor, Cambridge, Massachusetts 02142, USA~…}
}

@article{hartmann2019comparing,
  title={Comparing automated text classification methods},
  author={Hartmann, Jochen and Huppertz, Juliana and Schamp, Christina and Heitmann, Mark},
  journal={International Journal of Research in Marketing},
  volume={36},
  number={1},
  pages={20--38},
  year={2019},
  publisher={Elsevier}
}

@article{scalco2026detect,
  title={How to Detect Information Voids Using Longitudinal Data from Social Media and Web Searches},
  author={Scalco, Irene and Gesualdo, Francesco and Cerqueti, Roy and Cinelli, Matteo},
  journal={arXiv preprint arXiv:2602.15476},
  year={2026}
}

@article{mredula2022review,
  title={A review on the trends in event detection by analyzing social media platforms’ data},
  author={Mredula, Motahara Sabah and Dey, Noyon and Rahman, Md Sazzadur and Mahmud, Imtiaz and Cho, You-Ze},
  journal={Sensors},
  volume={22},
  number={12},
  pages={4531},
  year={2022},
  publisher={MDPI}
}

@article{wu2025people,
  title={Why people share misinformation on social media? An integration of affordance and flow theories},
  author={Wu, Manli and Wu, Tailai and Xiao, Yushan},
  journal={Humanities and Social Sciences Communications},
  volume={12},
  number={1},
  pages={1--11},
  year={2025},
  publisher={Palgrave}
}

@article{dongqi2026expanding,
  title={Expanding discourse and advocating stances: Social bots action strategies in carbon neutrality discussions},
  author={Dongqi, Yan and Wujiong, Ren and Junchen, Yao and Yuduo, Wu and Yuan, He and Zhang, Hongzhong},
  journal={Journal of Cleaner Production},
  volume={544},
  pages={147591},
  year={2026},
  publisher={Elsevier}
}

@article{arceneaux2026social,
  title={Social Bots as Agenda-Builders: Evaluating the Impact of Algorithmic Amplification on Organizational Messaging},
  author={Arceneaux, Phillip and Anderson, Joshua and Lukito, Josephine and Shah, Mansi and Kiousis, Spiro},
  journal={Journal of Public Relations Research},
  pages={1--34},
  year={2026},
  publisher={Taylor \& Francis}
}

@article{jacobs2024whatisdemocracy,
  title={\# WhatIsDemocracy: finding key actors in a Chinese influence campaign},
  author={Jacobs, Charity S and Carley, Kathleen M},
  journal={Computational and Mathematical Organization Theory},
  volume={30},
  number={2},
  pages={127--147},
  year={2024},
  publisher={Springer}
}

@article{phillips2025emotions,
  title={Emotions moderate the influence of moral values on attitude stability},
  author={Phillips, Samantha C and Ng, Lynnette Hui Xian and Zhou, Wenqi and Carley, Kathleen M},
  journal={Computational and Mathematical Organization Theory},
  volume={31},
  number={4},
  pages={277--298},
  year={2025},
  publisher={Springer}
}

@article{pennycook2021psychology,
  title={The psychology of fake news},
  author={Pennycook, Gordon and Rand, David G},
  journal={Trends in cognitive sciences},
  volume={25},
  number={5},
  pages={388--402},
  year={2021},
  publisher={Elsevier}
}

@article{alieva2026dynamics,
  title={Dynamics of Russian anti-war discourse on X (Twitter): a computational analysis using NLP and network methods},
  author={Alieva, Iuliia and Carley, Kathleen M},
  journal={npj Complexity},
  volume={3},
  number={1},
  pages={2},
  year={2026},
  publisher={Nature Publishing Group UK London}
}

@article{wu2025research,
  title={Research on the influence mechanism of emotional communication on Twitter (X) and the effect of spreading public anger},
  author={Wu, Chunqiong and Jiang, Shan and Sun, Jianhong and Liu, Yingqi},
  journal={Acta Psychologica},
  volume={260},
  pages={105560},
  year={2025},
  publisher={Elsevier}
}

@article{crawford2025blending,
  title={Blending emotion and logic in health messaging strategy: Audience perception of message appeals in anti-and pro-vaccination memes},
  author={Crawford, Elizabeth Crisp and Thomas, Laura E and Yakubova, Muhabbat and Anderson, Murphy},
  journal={Health Marketing Quarterly},
  volume={42},
  number={4},
  pages={422--436},
  year={2025},
  publisher={Taylor \& Francis}
}

@inproceedings{grimme2018changing,
  title={Changing perspectives: Is it sufficient to detect social bots?},
  author={Grimme, Christian and Assenmacher, Dennis and Adam, Lena},
  booktitle={Social Computing and Social Media. User Experience and Behavior: 10th International Conference, SCSM 2018, Held as Part of HCI International 2018, Las Vegas, NV, USA, July 15--20, 2018, Proceedings, Part I},
  pages={445--461},
  year={2018},
  organization={Springer}
}

@article{steinmetz2021liking,
  title={Liking, tweeting and posting: an analysis of community engagement through social media platforms},
  author={Steinmetz, Christine and Rahmat, Homa and Marshall, Nancy and Bishop, Kate and Thompson, Susan and Park, Miles and Corkery, Linda and Tietz, Christian},
  journal={Urban Policy and Research},
  volume={39},
  number={1},
  pages={85--105},
  year={2021},
  publisher={Taylor {\&} Francis}
}

@inproceedings{ng2025appeal,
  author={Ng, Lynnette Hui Xian and Zhou, Wenqi and Carley, Kathleen M.},
  title={Appeal and Scope of Misinformation Spread by AI Agents and Humans},
  booktitle={Proceedings of the Americas Conference on Information Systems (AMCIS)},
  year={2025},
  url={https://aisel.aisnet.org/amcis2025/social_comput/social_comput/6}
}

@inproceedings{ibrahim2015persuasive,
  title={Persuasive impact of online media: investigating the influence of visual persuasion},
  author={Ibrahim, Nurulhuda and Wong, Kok Wai and Shiratuddin, Mohd Fairuz},
  booktitle={2015 Asia Pacific Conference on Multimedia and Broadcasting},
  pages={1--7},
  year={2015},
  organization={IEEE}
}

@article{sun2024heuristic,
  title={Do Heuristic Cues Affect Misinformation Sharing? Evidence From a Meta-Analysis},
  author={Sun, Yanqing and Xie, Juan},
  journal={Journalism \& Mass Communication Quarterly},
  pages={10776990241284597},
  year={2024},
  publisher={SAGE Publications Sage CA: Los Angeles, CA}
}

@article{safadi2024effect,
  title={The Effect of Bots on Human Interaction in Online Communities.},
  author={Safadi, Hani and Lalor, John P and Berente, Nicholas},
  journal={MIS Quarterly},
  volume={48},
  number={3},
  year={2024}
}

@article{kim2018rumor,
  title={Rumor has it: The effects of virality metrics on rumor believability and transmission on Twitter},
  author={Kim, Ji Won},
  journal={New Media {\&} Society},
  volume={20},
  number={12},
  pages={4807--4825},
  year={2018},
  publisher={SAGE Publications}
}

@article{prollochs2023mechanisms,
  title={Mechanisms of true and false rumor sharing in social media: collective intelligence or herd behavior?},
  author={Pr{\"o}llochs, Nicolas and Feuerriegel, Stefan},
  journal={Proceedings of the ACM on Human-Computer Interaction},
  volume={7},
  number={CSCW2},
  pages={1--38},
  year={2023},
  publisher={ACM}
}

@article{chen2021persuasion,
  title={Persuasion strategies of misinformation-containing posts in the social media},
  author={Chen, Sijing and Xiao, Lu and Mao, Jin},
  journal={Information Processing {\&} Management},
  volume={58},
  number={5},
  pages={102665},
  year={2021},
  publisher={Elsevier}
}

@article{al2019viral,
  title={Viral news on social media},
  author={Al-Rawi, Ahmed},
  journal={Digital Journalism},
  volume={7},
  number={1},
  pages={63--79},
  year={2019},
  publisher={Taylor {\&} Francis}
}

@article{stella2018bots,
  title={Bots increase exposure to negative and inflammatory content in online social systems},
  author={Stella, Massimo and Ferrara, Emilio and De Domenico, Manlio},
  journal={Proceedings of the National Academy of Sciences},
  volume={115},
  number={49},
  pages={12435--12440},
  year={2018},
  publisher={National Academy of Sciences}
}

@misc{memon2020characterizing,
  title={Characterizing {COVID-19} misinformation communities using a novel Twitter dataset},
  author={Memon, Shahan Ali and Carley, Kathleen M.},
  year={2020},
  eprint={2008.00791},
  archivePrefix={arXiv},
  primaryClass={cs.SI}
}

@article{fu2020social,
  title={Social media overload, exhaustion, and use discontinuance: Examining the effects of information overload, system feature overload, and social overload},
  author={Fu, Shaoxiong and Li, Hongxiu and Liu, Yong and Pirkkalainen, Henri and Salo, Markus},
  journal={Information Processing {\&} Management},
  volume={57},
  number={6},
  pages={102307},
  year={2020},
  publisher={Elsevier}
}

@article{shao2018spread,
  title={The spread of low-credibility content by social bots},
  author={Shao, Chengcheng and Ciampaglia, Giovanni Luca and Varol, Onur and Yang, Kai-Cheng and Flammini, Alessandro and Menczer, Filippo},
  journal={Nature Communications},
  volume={9},
  number={1},
  pages={4787},
  year={2018},
  publisher={Nature Publishing Group}
}

@article{muhammed2022disaster,
  title={The disaster of misinformation: A review of research in social media},
  author={Muhammed T., Sadiq and Mathew, Saji K.},
  journal={International Journal of Data Science and Analytics},
  volume={13},
  number={4},
  pages={271--285},
  year={2022},
  publisher={Springer}
}

@article{petty2008persuasion,
  title={Persuasion: From single to multiple to metacognitive processes},
  author={Petty, Richard E. and Brinol, Pablo},
  journal={Perspectives on Psychological Science},
  volume={3},
  number={2},
  pages={137--147},
  year={2008},
  publisher={SAGE Publications}
}

@article{marwick2011tweet,
  title={I tweet honestly, I tweet passionately: Twitter users, context collapse, and the imagined audience},
  author={Marwick, Alice E. and Boyd, Danah},
  journal={New Media {\&} Society},
  volume={13},
  number={1},
  pages={114--133},
  year={2011},
  publisher={SAGE Publications}
}

@incollection{gallois2005communication,
  title={Communication accommodation theory},
  author={Gallois, Cindy and Ogay, Tania and Giles, Howard},
  booktitle={Theorizing about Intercultural Communication},
  pages={121--148},
  year={2005}
}

@article{min2009cultural,
  title={A cultural paradox in authority-based advertising},
  author={Min Jung, Jae and Polyorat, Kawpong and Kellaris, James J.},
  journal={International Marketing Review},
  volume={26},
  number={6},
  pages={601--632},
  year={2009},
  publisher={Emerald Group Publishing}
}

@article{pennebaker2003psychological,
  title={Psychological aspects of natural language use: Our words, our selves},
  author={Pennebaker, James W. and Mehl, Matthias R. and Niederhoffer, Kate G.},
  journal={Annual Review of Psychology},
  volume={54},
  number={1},
  pages={547--577},
  year={2003},
  publisher={Annual Reviews}
}

@article{ng2022pro,
  title={Pro or Anti? A social influence model of online stance flipping},
  author={Ng, Lynnette Hui Xian and Carley, Kathleen M.},
  journal={IEEE Transactions on Network Science and Engineering},
  volume={10},
  number={1},
  pages={3--19},
  year={2022},
  publisher={IEEE}
}

@article{knapp1944psychology,
  title={A psychology of rumor},
  author={Knapp, Robert H.},
  journal={Public Opinion Quarterly},
  volume={8},
  number={1},
  pages={22--37},
  year={1944},
  publisher={Oxford University Press}
}

@inproceedings{beskow2018bot,
  title={Bot-hunter: A tiered approach to detecting {\&} characterizing automated activity on Twitter},
  author={Beskow, David M. and Carley, Kathleen M.},
  booktitle={International Conference on Social Computing, Behavioral-Cultural Modeling and Prediction and Behavior Representation in Modeling and Simulation},
  year={2018}
}

@article{ng2022stabilizing,
  title={Stabilizing a supervised bot detection algorithm: How much data is needed for consistent predictions?},
  author={Ng, Lynnette Hui Xian and Robertson, Dawn C. and Carley, Kathleen M.},
  journal={Online Social Networks and Media},
  volume={28},
  pages={100198},
  year={2022},
  publisher={Elsevier}
}

@article{trent2022trust,
  title={Trust in government, intention to vaccinate and {COVID-19} vaccine hesitancy: A comparative survey of five large cities in the United States, United Kingdom, and Australia},
  author={Trent, Mallory and Seale, Holly and Chughtai, Abrar Ahmad and Salmon, Daniel and MacIntyre, C. Raina},
  journal={Vaccine},
  volume={40},
  number={17},
  pages={2498--2505},
  year={2022},
  publisher={Elsevier}
}

@inproceedings{varol2017online,
  title={Online human-bot interactions: Detection, estimation, and characterization},
  author={Varol, Onur and Ferrara, Emilio and Davis, Clayton and Menczer, Filippo and Flammini, Alessandro},
  booktitle={Proceedings of the International AAAI Conference on Web and Social Media},
  volume={11},
  number={1},
  pages={280--289},
  year={2017}
}

@article{menczer2020attention,
  title={The attention economy},
  author={Menczer, Filippo and Hills, Thomas},
  journal={Scientific American},
  volume={323},
  number={6},
  pages={54--61},
  year={2020}
}

@article{ng2022cross,
  title={Cross-platform information spread during the January 6th capitol riots},
  author={Ng, Lynnette Hui Xian and Cruickshank, Iain J and Carley, Kathleen M},
  journal={Social Network Analysis and Mining},
  volume={12},
  number={1},
  pages={133},
  year={2022},
  publisher={Springer}
}

@misc{twitterTwittersRecommendation,
  author={Twitter},
  title={Twitter's Recommendation Algorithm --- blog.twitter.com},
  year={2023},
  url={https://blog.twitter.com/engineering/en_us/topics/open-source/2023/twitter-recommendation-algorithm},
  note={{[Accessed 2024-01-19]}}
}

@inproceedings{zhang2023twhin,
  title={Twhin-bert: A socially-enriched pre-trained language model for multilingual tweet representations at Twitter},
  author={Zhang, Xinyang and Malkov, Yury and Florez, Omar and Park, Serim and McWilliams, Brian and Han, Jiawei and El-Kishky, Ahmed},
  booktitle={Proceedings of the 29th ACM SIGKDD Conference on Knowledge Discovery and Data Mining},
  pages={5597--5607},
  year={2023}
}

@misc{smith2016mean,
  title={Mean affective ratings of 929 identities, 814 behaviors, and 660 modifiers by University of Georgia and Duke University undergraduates and by community members in Durham, NC, in 2012--2014},
  author={Smith-Lovin, Lynn and Robinson, Dawn T. and Cannon, Bryan C. and Clark, Jesse K. and Freeland, Robert and Morgan, Jonathan H. and Rogers, Kimberly B.},
  year={2016},
  howpublished={University of Georgia: Distributed at UGA Affect Control Theory Website},
  url={http://research.franklin.uga.edu/act}
}

@article{tversky1974judgment,
  title={Judgment under Uncertainty: Heuristics and Biases},
  author={Tversky, Amos and Kahneman, Daniel},
  journal={Science},
  volume={185},
  number={4157},
  pages={1124--1131},
  year={1974},
  publisher={American Association for the Advancement of Science}
}

@article{slechten2022adapting,
  title={Adapting the selective exposure perspective to algorithmically governed platforms: The case of Google Search},
  author={Slechten, Laura and Courtois, C{\'e}dric and Coenen, Lennert and Zaman, Bieke},
  journal={Communication Research},
  volume={49},
  number={8},
  pages={1039--1065},
  year={2022},
  publisher={Sage Publications}
}

@article{xu2022characterizing,
  title={Characterizing the roles of bots on Twitter during the {COVID-19} infodemic},
  author={Xu, Wentao and Sasahara, Kazutoshi},
  journal={Journal of Computational Social Science},
  volume={5},
  number={1},
  pages={591--609},
  year={2022},
  publisher={Springer}
}

@article{cargnino2021interplay,
  title={The interplay of online network homogeneity, populist attitudes, and conspiratorial beliefs: Empirical evidence from a survey on German Facebook users},
  author={Cargnino, Manuel},
  journal={International Journal of Public Opinion Research},
  volume={33},
  number={2},
  pages={337--353},
  year={2021},
  publisher={Oxford University Press}
}

@inproceedings{herman2023page,
  author={Herman, L.},
  title={For who page? TikTok creators’ algorithmic dependencies},
  booktitle={IASDR 2023: Life-Changing Design},
  editor={De Sainz Molestina, D. and Galluzzo, L. and Rizzo, F. and Spallazzo, D.},
  year={2023},
  address={Milan, Italy},
  month=oct,
  note={October 9--13},
  doi={10.21606/iasdr.2023.576},
  url={https://doi.org/10.21606/iasdr.2023.576}
}

@misc{npr,
  author={NPR},
  title={One of the Most Influential Voices in Vaccine Misinformation Is a Doctor},
  year={2021},
  url={https://www.npr.org/2021/08/08/1025845675/one-of-the-most-influential-voices-in-vaccine-misinformation-is-a-doctor},
  note={{[Accessed 2024-02-11]}}
}

@article{lin2016social,
  title={Social media and credibility indicators: The effect of influence cues},
  author={Lin, Xialing and Spence, Patric R. and Lachlan, Kenneth A.},
  journal={Computers in Human Behavior},
  volume={63},
  pages={264--271},
  year={2016},
  publisher={Elsevier}
}

@article{harris2024perceived,
  title={Perceived experts are prevalent and influential within an antivaccine community on Twitter},
  author={Harris, Mallory J. and Murtfeldt, Ryan and Wang, Shufan and Mordecai, Erin A. and West, Jevin D.},
  journal={PNAS Nexus},
  volume={3},
  number={2},
  pages={pgae007},
  year={2024},
  publisher={Oxford University Press US}
}

@article{wawrzuta2021characteristics,
  title={Characteristics of antivaccine messages on social media: systematic review},
  author={Wawrzuta, Dominik and Jaworski, Mariusz and Gotlib, Joanna and Panczyk, Mariusz},
  journal={Journal of Medical Internet Research},
  volume={23},
  number={6},
  pages={e24564},
  year={2021},
  publisher={JMIR Publications}
}

@article{seckin2024mechanisms,
  title={Mechanisms Driving Online Vaccine Debate During the {COVID-19} Pandemic},
  author={Seckin, Ozgur Can and Atalay, Aybuke and Otenen, Ege and Duygu, Umut and Varol, Onur},
  journal={Social Media {\&} Society},
  volume={10},
  number={1},
  pages={20563051241229657},
  year={2024},
  publisher={SAGE Publications}
}

@inproceedings{jacobs2023tracking,
  title={Tracking China’s cross-strait bot networks against Taiwan},
  author={Jacobs, Charity S. and Ng, Lynnette Hui Xian and Carley, Kathleen M.},
  booktitle={International Conference on Social Computing, Behavioral-Cultural Modeling and Prediction and Behavior Representation in Modeling and Simulation},
  pages={115--125},
  year={2023},
  organization={Springer}
}

@article{lee2024correcting,
  title={Correcting vaccine misinformation on social media: the inadvertent effects of repeating misinformation within such corrections on {COVID-19} vaccine misperceptions},
  author={Lee, Jiyoung and Bissell, Kim},
  journal={Current Psychology},
  pages={1--13},
  year={2024},
  publisher={Springer}
}

@article{danaditya2022curious,
  title={From curious hashtags to polarized effect: profiling coordinated actions in Indonesian Twitter discourse},
  author={Danaditya, Adya and Ng, Lynnette Hui Xian and Carley, Kathleen M.},
  journal={Social Network Analysis and Mining},
  volume={12},
  number={1},
  pages={105},
  year={2022},
  publisher={Springer}
}

@misc{lin_savio_huang_steiger_guevara_szostak_pennycook_rand_2024,
  title={Accuracy prompts protect professional content moderators from the illusory truth effect},
  url={https://osf.io/preprints/psyarxiv/gswm6},
  doi={10.31234/osf.io/gswm6},
  publisher={PsyArXiv},
  author={Lin, Hause and Savio, Marlyn T and Huang, Xieyining and Steiger, Miriah and Guevara, Rachel and Szostak, Dali and Pennycook, Gordon and Rand, David G},
  year={2024},
  month=mar
}

@inproceedings{mazza2019rtbust,
  title={Rtbust: Exploiting temporal patterns for botnet detection on Twitter},
  author={Mazza, Michele and Cresci, Stefano and Avvenuti, Marco and Quattrociocchi, Walter and Tesconi, Maurizio},
  booktitle={Proceedings of the 10th ACM Conference on Web Science},
  pages={183--192},
  year={2019}
}

@inproceedings{wang2020case,
  title={A case study in Twitter bot identification: Are they still a problem?},
  author={Wang, Tiange and Wu, Fengkai and Sinnott, Richard O.},
  booktitle={2020 Seventh International Conference on Social Networks Analysis, Management and Security (SNAMS)},
  pages={1--8},
  year={2020},
  organization={IEEE}
}

@article{yuan2019examining,
  title={Examining emergent communities and social bots within the polarized online vaccination debate in Twitter},
  author={Yuan, Xiaoyi and Schuchard, Ross J. and Crooks, Andrew T.},
  journal={Social Media {\&} Society},
  volume={5},
  number={3},
  pages={2056305119865465},
  year={2019},
  publisher={SAGE Publications}
}

@article{guo2022does,
  title={How does multi-platform social media use lead to biased news engagement? Examining the role of counter-attitudinal incidental exposure, cognitive elaboration, and network homogeneity},
  author={Guo, Jing and Chen, Hsuan-Ting},
  journal={Social Media {\&} Society},
  volume={8},
  number={4},
  pages={20563051221129140},
  year={2022},
  publisher={SAGE Publications}
}

@book{sunstein2018republic,
  title={\#Republic: Divided Democracy in the Age of Social Media},
  author={Sunstein, Cass},
  year={2018},
  publisher={Princeton University Press}
}

@article{vosoughi2018spread,
  title={The spread of true and false news online},
  author={Vosoughi, Soroush and Roy, Deb and Aral, Sinan},
  journal={Science},
  volume={359},
  number={6380},
  pages={1146--1151},
  year={2018},
  publisher={American Association for the Advancement of Science}
}

@article{modgil2021confirmation,
  title={A confirmation bias view on social media induced polarisation during {COVID-19}},
  author={Modgil, Sachin and Singh, Rohit Kumar and Gupta, Shivam and Dennehy, Denis},
  journal={Information Systems Frontiers},
  pages={1--25},
  year={2021},
  publisher={Springer}
}

@misc{li2024social,
  title={Social bots sour activist sentiment without eroding engagement},
  author={Li, Linda and Vasarhelyi, Orsolya and Vedres, Balazs},
  year={2024},
  eprint={2403.12904},
  archivePrefix={arXiv},
  primaryClass={cs.SI}
}

@inproceedings{ng2021bot,
  title={Bot-based emotion behavior differences in images during {Kashmir Black Day} event},
  author={Ng, Lynnette Hui Xian and Carley, Kathleen M.},
  booktitle={International Conference on Social Computing, Behavioral-Cultural Modeling and Prediction and Behavior Representation in Modeling and Simulation},
  pages={184--194},
  year={2021},
  organization={Springer}
}

@misc{gnetresearchPastelQAnon,
  author={Argentino, Marc-Andr{\'e}},
  title={Pastel QAnon -- GNET},
  year={2021},
  url={https://gnet-research.org/2021/03/17/pastel-qanon/},
  note={{[Accessed 2024-02-12]}}
}

@article{metzger2013credibility,
  title={Credibility and trust of information in online environments: The use of cognitive heuristics},
  author={Metzger, Miriam J. and Flanagin, Andrew J.},
  journal={Journal of Pragmatics},
  volume={59},
  pages={210--220},
  year={2013},
  publisher={Elsevier}
}

@article{pennycook2019fighting,
  title={Fighting misinformation on social media using crowdsourced judgments of news source quality},
  author={Pennycook, Gordon and Rand, David G.},
  journal={Proceedings of the National Academy of Sciences},
  volume={116},
  number={7},
  pages={2521--2526},
  year={2019},
  publisher={National Academy of Sciences}
}

@article{kenny2024duped,
  title={Duped by bots: why some are better than others at detecting fake social media personas},
  author={Kenny, Ryan and Fischhoff, Baruch and Davis, Alex and Carley, Kathleen M. and Canfield, Casey},
  journal={Human Factors},
  volume={66},
  number={1},
  pages={88--102},
  year={2024},
  publisher={SAGE Publications}
}

@article{acerbi2019cognitive,
  title={Cognitive attraction and online misinformation},
  author={Acerbi, Alberto},
  journal={Palgrave Communications},
  volume={5},
  number={1},
  year={2019},
  publisher={Springer Nature}
}

@misc{tan2023botpercent,
  title={BotPercent: Estimating Twitter bot populations from groups to crowds},
  author={Tan, Zhaoxuan and Feng, Shangbin and Sclar, Melanie and Wan, Herun and Luo, Minnan and Choi, Yejin and Tsvetkov, Yulia},
  year={2023},
  eprint={2302.00381},
  archivePrefix={arXiv},
  primaryClass={cs.SI}
}

@article{fukuda2022estimating,
  title={Estimating the bot population on Twitter via random walk based sampling},
  author={Fukuda, Mei and Nakajima, Kazuki and Shudo, Kazuyuki},
  journal={IEEE Access},
  volume={10},
  pages={17201--17211},
  year={2022},
  publisher={IEEE}
}

@article{chen2022social,
  title={Social network behavior and public opinion manipulation},
  author={Chen, Long and Chen, Jianguo and Xia, Chunhe},
  journal={Journal of Information Security and Applications},
  volume={64},
  pages={103060},
  year={2022},
  publisher={Elsevier}
}

@incollection{ndlela2020social,
  title={Social media algorithms, bots and elections in Africa},
  author={Ndlela, Martin N.},
  booktitle={Social Media and Elections in Africa, Volume 1: Theoretical Perspectives and Election Campaigns},
  pages={13--37},
  year={2020},
  publisher={Springer}
}

@article{himelein2021bots,
  title={Bots and misinformation spread on social media: Implications for {COVID-19}},
  author={Himelein-Wachowiak, McKenzie and Giorgi, Salvatore and Devoto, Amanda and Rahman, Muhammad and Ungar, Lyle and Schwartz, H. Andrew and Epstein, David H. and Leggio, Lorenzo and Curtis, Brenda},
  journal={Journal of Medical Internet Research},
  volume={23},
  number={5},
  pages={e26933},
  year={2021},
  publisher={JMIR Publications}
}

@article{broniatowski2018weaponized,
  title={Weaponized health communication: Twitter bots and Russian trolls amplify the vaccine debate},
  author={Broniatowski, David A. and Jamison, Amelia M. and Qi, SiHua and AlKulaib, Lulwah and Chen, Tao and Benton, Adrian and Quinn, Sandra C. and Dredze, Mark},
  journal={American Journal of Public Health},
  volume={108},
  number={10},
  pages={1378--1384},
  year={2018},
  publisher={American Public Health Association}
}

@article{braca2023developing,
  title={Developing persuasive systems for marketing: the interplay of persuasion techniques, customer traits and persuasive message design},
  author={Braca, Annye and Dondio, Pierpaolo},
  journal={Italian Journal of Marketing},
  volume={2023},
  number={3},
  pages={369--412},
  year={2023},
  publisher={Springer}
}

@inproceedings{boyd2010tweet,
  title={Tweet, tweet, retweet: Conversational aspects of retweeting on Twitter},
  author={Boyd, Danah and Golder, Scott and Lotan, Gilad},
  booktitle={2010 43rd Hawaii International Conference on System Sciences},
  pages={1--10},
  year={2010},
  organization={IEEE}
}

@article{horner2021emotions,
  title={Emotions: The unexplored fuel of fake news on social media},
  author={Horner, Christy Galletta and Galletta, Dennis and Crawford, Jennifer and Shirsat, Abhijeet},
  journal={Journal of Management Information Systems},
  volume={38},
  number={4},
  pages={1039--1066},
  year={2021},
  publisher={Taylor {\&} Francis}
}

@article{tausczik2010psychological,
  title={The psychological meaning of words: {LIWC} and computerized text analysis methods},
  author={Tausczik, Yla R. and Pennebaker, James W.},
  journal={Journal of Language and Social Psychology},
  volume={29},
  number={1},
  pages={24--54},
  year={2010},
  publisher={Sage Publications}
}

@article{mocanu2015collective,
  title={Collective attention in the age of (mis) information},
  author={Mocanu, Delia and Rossi, Luca and Zhang, Qian and Karsai, Marton and Quattrociocchi, Walter},
  journal={Computers in Human Behavior},
  volume={51},
  pages={1198--1204},
  year={2015},
  publisher={Elsevier}
}

@article{zhao2020promoting,
  title={Promoting users’ intention to share online health articles on social media: The role of confirmation bias},
  author={Zhao, Haiping and Fu, Shaoxiong and Chen, Xiaoyu},
  journal={Information Processing {\&} Management},
  volume={57},
  number={6},
  pages={102354},
  year={2020},
  publisher={Elsevier}
}

@article{ng2021coronavirus,
  title={“The coronavirus is a bioweapon”: classifying coronavirus stories on fact-checking sites},
  author={Ng, Lynnette Hui Xian and Carley, Kathleen M},
  journal={Computational and Mathematical Organization Theory},
  volume={27},
  number={2},
  pages={179--194},
  year={2021},
  publisher={Springer}
}

@book{friedkin2011social,
  title={Social influence network theory: A sociological examination of small group dynamics},
  author={Friedkin, Noah E and Johnsen, Eugene C},
  volume={33},
  year={2011},
  publisher={Cambridge University Press}
}

@article{chaiken1980heuristic,
  title={Heuristic versus systematic information processing and the use of source versus message cues in persuasion.},
  author={Chaiken, Shelly},
  journal={Journal of personality and social psychology},
  volume={39},
  number={5},
  pages={752},
  year={1980},
  publisher={American Psychological Association}
}

@article{rathje2025psychology,
  title={The psychology of virality},
  author={Rathje, Steve and Van Bavel, Jay J},
  journal={Trends in Cognitive Sciences},
  year={2025},
  publisher={Elsevier}
}

@article{evans2013dual,
  title={Dual-process theories of higher cognition: Advancing the debate},
  author={Evans, Jonathan St BT and Stanovich, Keith E},
  journal={Perspectives on psychological science},
  volume={8},
  number={3},
  pages={223--241},
  year={2013},
  publisher={Sage Publications Sage CA: Los Angeles, CA}
}

\section*{Funding}
The research for this paper was supported by the following grants: Cognizant Center of Excellence Content Moderation Research Program, Office of Naval Research (Bothunter, N000141812108) and Scalable Technologies for Social Cybersecurity/ARMY (W911NF20D0002). The views and conclusions are those of the authors and should not be interpreted as representing the official policies, either expressed or implied.

\section*{Author contributions statement}
L.H.X.N. (Conceptualization, Data curation, Methodology, Analysis, Writing); W. Z. (Conceptualization, Methodology, Analysis, Writing); K.M.C. (Conceptualization, Funding acquistion, Project administration, Writing). All authors reviewed the manuscript. 

\section*{Competing Interests} The authors declare that there is no competing interests. 

\section*{Data Availability Statement} 
The COVID-19 dataset used in this study was obtained from the the paper ``A global comparison of social media bot and human characteristics", that was published in Scientific Reports in 2025 \cite{ng2025global}. Since this is a dataset containing social media platform data, the data cannot be publicly available. However, one can contact the authors to request to request the data and code from this study, in accordance to the data sharing policies of the social media platforms.

\clearpage
\appendix

\renewcommand{\thefigure}{S\arabic{figure}}
\renewcommand{\thetable}{S\arabic{table}}
\renewcommand{\thesection}{S\arabic{section}}

\setcounter{figure}{0}
\setcounter{table}{0}
\setcounter{section}{0}

\section*{Supplementary Material}
\section{Dataset Construction}
\label{supp:dataset_construction}
We constructed the foundational dataset using the X Developer API V1 streaming function, collecting a 1\% sample of tweets from July 2020 to July 2021. The specific set of keywords used to collect the data was: COVID-19, coronavirus, wuhan virus, wuhanvirus, 2019nCoV, NCoV, vaccine, vax, and vaccination. This dataset is termed as the COVID-19  Dataset. 

\textbf{Ethical Statement.} We extracted only publicly available data from the social media platform and made no attempt to retrieve protected posts. We did not process or expose unique personal identifiers of the social media accounts during our analysis.

\subsection{Extracting Tweets} To isolate contested and unverified narratives, we filtered the corpus (``COVID-19 Dataset") against an expertly annotated reference dataset \cite{memon2020characterizing}. This reference dataset contains manual categorizations of prominent COVID-19 narratives (e.g., fake cure, fake fact) that stemmed from the 2020 COVID-19 pandemic \cite{ng2021coronavirus}.

Prior to vectorization, we pre-processed the tweets in COVID-19 Dataset by removing artifacts in COVID-19 Dataset such as URLs and user mentions. We then generated dense vector representations of the tweet text using TwHIN-BERT embeddings \cite{zhang2023twhin}. Because this embedding architecture was trained by the platform's research team and utilized in its front-feed recommendation algorithm\cite{twitterTwittersRecommendation}, it is highly suitable for representing our specific social media data.

To validate these text embedding representations, we trained a multi-class logistic classifier on a randomly selected 80\% subset of the reference data, using the embeddings as inputs and the narrative categories \ (e.g., fake cure, fake fact etc) as outputs. Testing the classifier on the remaining 20\% holdout set yielded an accuracy of 50.13\%, a reasonably high score that significantly exceeds the random-chance baseline of 20\%. The random-chance baseline is a common mathematical baseline, where classification performance is assessed by how much it deviates from the chance level of $1/k$ in a $k$-class classification problem\cite{hastie2009elements,combrisson2015exceeding}.

We then classified the unannotated tweets in COVID-19 Dataset using a cosine similarity procedure adapted from prior matching studies \cite{ng2022cross,ng2025appeal}. Unannotated tweet vectors with a cosine distance of $\geq 0.8$ relative to a reference vector (indicating 80\% semantic similarity) were assigned that corresponding narrative class. We operated under the assumption that these retrieved tweets were true positives based on their close structural resemblance to the manually curated reference data. \autoref{tab:eg_humans_misinfo} and \autoref{tab:eg_bots_misinfo}  present examples of these matched tweets by humans and bots, respectively.

\begin{table}[h]
    \centering
    \begin{tabular}{|p{3cm}|p{10cm}|}
        \hline
        \textbf{Category} & \textbf{Example Tweet} \\ \hline
        Conspiracy & this is too perfect a bioweapon to have occurred naturally \newline\newline the liberals controlled msm and has created the hysteria around the coronavirus which was almost certainly created as bioweapon \\ \hline
        Fake cure & we need to get the word out about all these three hydroxychloroquine antibiotics and zic for the cure to covid \newline\newline hydroxychloroquine is the cure the government is trying to hide with vaccine \\ \hline
        Fake fact & if you mix 50 bleach \& 50 water \& dip a cleaning cloth or paper towel into it it's just as good as an anti micro for covid \\ \hline
        Fake treatment & sesame oil kills the new coronavirus \newline\newline anyone dealing with covid highly recommend a tea made of ginger garlic red onion lemon cinnamon amp honey helps the immune system  \\ \hline
        False public health responses & italy allows malaria and hiv drugs for covid19 treatment \newline\newline earlier this week india had sent to israel a 5 tonne cargo of medicines including anti malarial drug \\ \hline
    \end{tabular}
    \caption{\textbf{Examples of Low-Credibility Tweets by Humans}}
    \label{tab:eg_humans_misinfo}
\end{table}

\begin{table}[h]
    \centering
    \begin{tabular}{|p{6cm}|p{8cm}|}
        \hline
        \textbf{Category} & \textbf{Example Tweet} \\ \hline
        Conspiracy & This doctor shows evidence of how our governments are using COVID \& the subsequent vaccine as a genocidal weapon to kill us \\ \hline
        Fake cure & salt solution can cure covid19 \\ \hline
        Fake fact & Russia says no to booze after vaccine shot!! \newline\newline Man paralyzed after covid vaccine \\ \hline
        Fake treatment & Immediately add immunity-building/anti-inflammatory/anti-viral garlic/Vitamin D3 to the treatment mix! Garlic cuts colds by 50\% (COVID-19 is a form of a cold) \&; per the Israelis, 2000-5000 IUs of daily D3 cuts COVID+ cases 50\% also. Many people are stuck on vaccines/I'm 4 FOODS! \\ \hline
        False public health responses & The purpose of 'track \& trace' is for governments to use the manufactured 'Corona crisis' to install Orwellian Police State controls.  The same with the fake 'Corona vaccines' - which are not 'vaccines' at all.  Take them - or you will be unable to fly, etc. \\ \hline
    \end{tabular}
    \caption{\textbf{Examples of Low-Credibility Tweets by Bots}}
    \label{tab:eg_bots_misinfo}
\end{table}

\subsection{Bot User Annotation}
To classify account types, we used BotHunter, a tier-based bot detection model. BotHunter consists of several Random Forest classifiers trained on manually annotated social media data in a hierarchical fashion. The algorithm evaluates accounts across features such as user name, post texts, and user metadata (e.g., number of followers, number of likes), and returns a probability that the user is likely to be a bot. In standard benchmark evaluations, this model achieves robust classification accuracy of over 90\% \cite{beskow2018bot}. This algorithm has been used in the study of a wide variety of events, including war discourse\cite{alieva2026dynamics}, health pandemics\cite{phillips2025emotions}, and elections\cite{jacobs2024whatisdemocracy}.

The model assigned each user a probability score $P(bot)$ between 0 and 1. To prioritize precision, users with a probability score of $P(bot)\geq 0.70$ were labeled as automated accounts (bots). This threshold was referenced from previous work that performed a large-scale statistical analysis of BotHunter scores over time \cite{ng2022stabilizing}. Under this configuration, 519,337 accounts (23.1\%) were classified as bots. The remaining 1,738,645 accounts were classified as humans.

\section{Human Annotation and Algorithm Validation}
\label{sec:human_annotation}
Two domain experts manually annotated a random sample of n=800 the bot-authored tweets. These two domain experts were part of the research group of this study. As such, no external human subjects were involved in this process, and separate consent documentation was not required.

We sampled from the automated account subset because these tweets possess a higher baseline representation of bias triggers. We confirmed that the drawn sample covered all evaluated biases. The annotators were trained in the understanding of cognitive biases and had not seen the dataset prior to the task.

During annotation, a single tweet could be labeled with multiple bias types. If no bias trigger was found, the tweet remained unlabeled. The annotators were also asked to note down their evaluation process, which we subsequently codified as our pattern-matching computational heuristics. A third expert was recruited to resolve any disagreements.

Cognitive Dissonance, which requires longitudinal social network interactions, was excluded from this manual annotation. Its detection heuristics were instead formed using the Friedkin-Johnson social influence model \cite{friedkin2011social}, which has been validated in prior social media studies \cite{ng2022pro}. As detailed in the main text, all quantitative thresholds (e.g., $k \geq 3$) were mathematically defined using the distribution ceiling $\text{ceiling}(\mu+\sigma)$.

The two independent annotators reached an agreement 71.89\% of the time. Overall, the automated computational algorithms matched the human consensus with classification accuracy ranging from 50\% to 83\%, exceeding the random-assignment baseline probability of 14.28\%. A random-assignment baseline assumes a uniform prior over each of the cognitive bias labels. That is, it guesses each class with an equal probability. This chance-level baseline is commonly used in multiclass classification to contextualize the performance relative to random guesses \cite{hartmann2019comparing,opitz2024closer}. \autoref{tab:human_annotation_accuracy} summarizes the agreement percentage between the first two annotators and the computational accuracy of the developed heuristics.

For Authority Bias, we initially identified two distinct heuristics: one using explicit authorities (e.g., government officials, public figures, authoritative occupations like teachers or police) and another using implicit authorities (e.g., social media influencers, context-specific authorities). Although combining both implicit and explicit markers yields a higher overall accuracy ($\sim62\%$), we elected to use only the explicit formulation to ensure the methodology remains generalizable and reproducible without requiring a manually compiled list of influencers.

\begin{table}
    \centering
    \begin{tabular}{|p{5cm}|p{3cm}|p{3cm}|p{3cm}|}
    \hline
       \textbf{Bias} & Inter-Annotator Agreement & Algorithm Accuracy (\%) \\
    \hline
    Homophily Bias & 98.88 & 50.00 \\ \hline 
    Authority Bias (Implicit Authority) & 60.02 & 62.00 \\ \hline 
    Authority Bias (Explicit Authority) & 60.02 & 62.10 \\ \hline 
    Affect/ Negativity Bias & 69.00 & 54.07 \\ \hline 
    Illusory Truth Effect & 97.48 & 66.67 \\ \hline 
    Availability Bias & 67.46 & 82.93 \\ \hline 
    Confirmation Bias & 32.54 & 61.00\\ \hline 
    \end{tabular}
    \caption{\textbf{Annotation Statistics.} Inter-annotator agreement represents the proportion of tweets the first two annotators agreed upon. A score closer to 1 represents complete agreement, and a score closer to 0 represents complete disagreement. Accuracy (\%) denotes the proportion of annotated tweets (n=800) that our computational algorithm matched the human annotation. }
    \label{tab:human_annotation_accuracy}
\end{table}

\section{Measuring Presence of Bias Triggers}
This section details the empirical and statistical foundations for our measurement of computational bias triggers. The operationalization of these triggers is grounded in a comprehensive survey of prior empirical literature observing cognitive heuristics in social media environments, detailed in  \autoref{tab:bias_literature}. 
Furthermore, to ensure our computational thresholds (e.g., $k\geq3$ repetitions or emotional words) were not arbitrary, we derived them mathematically from the baseline distributions within our dataset. \autoref{tab:desc_stats} provides the mean ($\mu$) and standard deviation ($\sigma$) for key structural and linguistic metrics across both automated and human-authored tweets. The quantitative thresholds utilized in our detection algorithms represent the ceiling values of these combined baseline distributions ($\text{ceiling}(\mu+\sigma)$), allowing us to systematically identify deliberate structural embeddings rather than baseline conversational noise.

\newgeometry{margin=1cm} 
\begin{landscape}

\begin{longtable}{|p{4cm}|p{20cm}|}
    \hline
    \textbf{Bias} & \textbf{Observations in Literature}\\ \hline
    \endfirsthead

    \multicolumn{2}{c}{\tablename\ \thetable{} -- continued from previous page}\\
    \hline
    \textbf{Bias} & \textbf{Observations in Literature}\\ \hline
    \endhead

    \hline
    \multicolumn{2}{r}{Continued on next page}\\
    \endfoot

    \hline
    \caption{\textbf{Empirical Observations of Cognitive Heuristics on Social Media in Literature.}}
    \label{tab:bias_literature}
    \endlastfoot

    \multicolumn{2}{|l|}{\textbf{Representativeness Heuristics}} \\ \hline
    Homophily Bias & Users preferentially retweet posts of similar political leanings \cite{xu2022characterizing}\\ \hline
    ~ & "Social network homogeneity": users are exposed to like-minded information within online networks \cite{cargnino2021interplay}\\ \hline
    ~ & TikTok's recommendation system that recommends videos based on match user demographics \cite{herman2023page}\\ \hline
    ~ & Homophily in social networks lead the mass audience to see more social media posts from the politicians that they align with \cite{haselswerdt2024echo} \\ \hline
    ~ & TikTok has clusters of politically homogeneous networks called ``political echo chambers", where users are exposed to political videos that are consistent with their attitudes \cite{li2025tiktok} \\ \hline

    Authority Bias & Tweets that tag influencers or important politicians lend credibility and enhance believability \cite{xu2022characterizing}\\ \hline
    ~ & A small but vocal minority of anti-COVID-19-vaccine medical professionals leverage their professional titles and medical expertise as evidence to persuade people of anti-vaccine narratives \cite{npr}\\ \hline
    ~ & Authority cues are most effective at inducing credibility bias \cite{lin2016social} \\ \hline
    ~ & Perceived experts are influential in spreading anti-vaccine information on social media \cite{harris2024perceived} \\ \hline
    ~ & Biomedical credentials raise perceived expertise of the user on X on the topic of COVID-19 vaccines \cite{jalbert2025perceived} \\ \hline

    \multicolumn{2}{|l|}{\textbf{Availability Heuristics}}\\ \hline
    Affect Bias & Antivaccine content tend to use more emotional framing rather than factual narratives, often expressing sentiments through images and videos to increase affective appeal \cite{wawrzuta2021characteristics}\\ \hline
    ~ & Aesthetically pleasing Instagram posts are used to spread QAnon-related content and conspiracy theories \cite{gnetresearchPastelQAnon}\\ \hline
    ~ & Automated accounts invoke emotions to drive calls to action during crises \cite{ng2021bot}\\ \hline
    ~ & Emotional messaging in visual memes increases persuasive impact for public health marketing \cite{crawford2025blending} \\ \hline
    
    Negativity Bias & Automated accounts increase audience exposure to negative/ inflammatory contents; they generate specific content with negative connotation that targets most influential individuals \cite{stella2018bots}\\ \hline
    ~ & Anti-vaccination groups use significantly more negative affect terms and references to death on Twitter \cite{seckin2024mechanisms}\\ \hline
    ~ & Automated accounts consistently display significantly more negative sentiment and demonstrate consistently negative impact during heated online periods \cite{li2024social}\\ \hline
    ~ & Anger-driven content on X have a disproportionately higher engagement and diffusion \cite{wu2025research} \\ \hline 
    
    Illusory Truth Effect & Automated accounts repeat almost identical phrases in messages on democracy ideals \cite{jacobs2023tracking}\\ \hline
    ~ & High-frequency repetition of identical message, even when paired with corrections, increase subsequent belief in unverified information \cite{lee2024correcting}\\ \hline
    ~ & Automated accounts repeat slight variation of a single message template in discussing religious issues surrounding Indonesia on Twitter \cite{danaditya2022curious}.\\ \hline
    ~ & Content moderators of COVID-19 headlines can be susceptible to illusory truth effect after repeated exposure to identical unverified news \cite{lin_savio_huang_steiger_guevara_szostak_pennycook_rand_2024}\\ \hline
    Availability Bias & The average number of retweets for low-credibility automated accounts is significantly higher than that of high-credibility users \cite{xu2022characterizing}\\ \hline
    ~ & High-frequency retweet behavior is a key feature used to identical social media automated accounts \cite{mazza2019rtbust,wang2020case}\\ \hline
    ~ & Automated accounts display hyper-social tendencies by initiating retweets in the coronavirus discourse \cite{yuan2019examining}\\ \hline

    \multicolumn{2}{|l|}{\textbf{Anchoring Heuristics}} \\ \hline
    Cognitive Dissonance & Peer pressure orchestrated by coordinated bots results in users changing their expressed stance towards the vaccine \cite{ng2022pro}\\ \hline
    & Exposure to counter-attitudinal information often causes people to anchor more firmly to their prior opinions \cite{guo2022does}\\ \hline
    Confirmation Bias & Engagement algorithms aggregate homogeneous content to reinforce users' existing beliefs, leading to the formation of echo chambers that marginalize opposing minority perspectives \cite{guo2022does,sunstein2018republic}\\ \hline
    ~ & Confirmation Bias on Twitter induced polarization and echo chamber formation during the COVID-19 pandemic \cite{modgil2021confirmation}\\ \hline
    & "Online selective exposure'': users maintain their prior belief despite exposure to diverse viewpoints in Google search engine results \cite{slechten2022adapting}\\ \hline

\end{longtable}

\end{landscape}
\restoregeometry

\begin{table}[h]
    \centering
    \begin{tabular}{|p{4cm}|p{1.5cm}|p{1.5cm}|p{1.5cm}|p{1.5cm}|p{1.5cm}|p{1.5cm}|}
    \hline
       ~ & \multicolumn{3}{|c|}{\textbf{Automated Accounts (bots)}} & \multicolumn{3}{|c|}{\textbf{Humans}} \\ \hline 
       \textbf{Metric} & \textbf{Mean} & \textbf{Median} & \textbf{Std Dev.} & \textbf{Mean} & \textbf{Median} & \textbf{Std Dev.} \\ \hline 
        Number of Favorites per tweet & 26.64 & 4.00 & 3036.38 & 0.34 & 0.00 & 140.54 \\ \hline 
        Number of  Quotes per tweets & 1.32 & 0.00 & 250.02 & 0.03 & 0.00 & 10.74 \\ \hline 
        Number of  Retweets per tweet & 8.65 & 3.00 & 603.91 & 0.97 & 0.00 & 43.18 \\ \hline 
        Number of  Replies per tweet & 2.29 & 0.00 & 143.48 & 0.02 & 0.00 & 9.42 \\ \hline 
        Number of  emotion words per tweet& 2.01 & 2.00 & 0.82 & 0.46 & 0.00 & 0.83 \\ \hline 
        Number of  negative words per tweet& 1.20 & 1.00 & 0.53 & 0.02 & 0.00 & 0.20 \\ \hline 
        Average number of tweets per user & 10.04 & 5.00 & 17.33 & 6.40 & 4.00 & 9.34 \\ \hline 
        Average number of retweets per user & 7.04 & 4.00 & 14.88  & 4.02 & 3.00 & 5.23 \\ \hline 
        Average number of engagement per tweet & 38.91 & 3.00 & 3927.65 & 1.37 & 0.00 & 182.45 \\ \hline 
    \end{tabular}
    \caption{\textbf{Descriptive Statistics of the Dataset}} Values inform the mathematical thresholds for computational trigger detection
    \label{tab:desc_stats}
\end{table}

\section{Regression Analysis between Bias Triggers and Engagement}
\autoref{tab:regression} presents the results of the regression analysis between the presence of bias triggers and tweet engagement, using the equation \autoref{eq:ols_supp}. One equation was constructed for bots, and another one for humans.

\begin{equation}
    \begin{split}
        log(Y^{k}) = \, & \alpha_0^{k} \\
        & + \alpha_1^{k}(\text{Homophily Bias}) \\
        & + \alpha_2^{k}(\text{Authority Bias}) \\
        & + \alpha_3^{k}(\text{Availability Bias}) \\
        & + \alpha_4^{k}(\text{Illusory Truth Effect}) \\
        & + \alpha_5^{k}(\text{Affect Bias}) \\
        & + \alpha_6^{k}(\text{Cognitive Dissonance}) \\
        & + \alpha_7^{k}(\text{Confirmation Bias}) \\
        & + \epsilon
    \end{split}
    \label{eq:ols_supp}
\end{equation}
where $k\in\{\text{Retweet Count, Quote Count, Favorite Count, Reply Count}\}$

\newgeometry{margin=1cm} 
\begin{landscape}

\begin{table}
    \centering
    \begin{tabular}{|c|p{2cm}|p{2cm}|p{2cm}|p{2cm}|p{2cm}|p{2cm}|p{2cm}|p{2cm}|}
    \hline
        ~ & \multicolumn{4}{|c|}{\textbf{Bots}} & \multicolumn{4}{|c|}{\textbf{Humans}} \\ \hline 
        \textbf{Bias} & \textbf{Retweet Count} & \textbf{Quote Count} & \textbf{Favorite Count} & \textbf{Reply Count} & \textbf{Retweet Count} & \textbf{Quote Count} & \textbf{Favorite Count} & \textbf{Reply Count} \\ \hline
        
        \textbf{Constant} & 0.32*** \newline (8.52E-4) & 0.07*** \newline (4.08E-4) & 0.27*** \newline (8.34E-4) & 0.09*** \newline (4.66E-4) & 0.03*** \newline (3.61E-4) & 1.26E-4*** \newline (3.02E-5) & 2.44E-4*** \newline (4.70E-5) & 1.17E-4*** \newline (2.99E-5) \\ \hline
        
        \textbf{Homophily Bias} & \textbf{-0.06}*** \newline (1.27E-2) &  -0.0003 \newline (6.07E-3)&  \-0.01 \newline (1.24E-2)&  -0.001 \newline (6.93E-3) & -0.008*** \newline (6.79E-3) &  -3.25E-4*** \newline (5.68E-4) &  -5.76E-4*** \newline (8.85E-4) &  -2.95E-4*** \newline 5.65E-4  \\ \hline
        
        \textbf{Authority Bias} & \textbf{-0.29}*** \newline (1.95E-2) & -\textbf{0.06}*** \newline (9.31E-3) &  \textbf{-0.25}*** \newline (1.90E-2) &  \textbf{-0.08}*** \newline (1.06E-2) &  \textbf{0.05}*** \newline (1.69E-3) &  1.16E-3*** \newline (1.41E-4) &  1.69E-3*** \newline (2.20E-4) &  1.12E-3*** \newline (1.41E-4)  \\ \hline
        
        \textbf{Availability Bias} & \textbf{-0.39}*** \newline (1.12E-3) &  \textbf{-0.08}*** \newline (5.34E-4) &  \textbf{-0.32}*** \newline (1.09E-3) &  \textbf{-0.11}*** \newline (6.09E-4) &  \textbf{-0.01}*** \newline (7.37E-4) & 4.43E-5*** \newline (6.17E-5) & 4.81E-5*** \newline (9.60E-5) & 4.31E-5*** \newline (1.41E-4)\\ \hline
        
        \textbf{Illusory Truth Effect} & \textbf{0.02}*** \newline (1.94E-3) &  \textbf{-0.02}*** \newline (9.30E-4) &  \textbf{-0.05}*** \newline (1.80E-3) &  \textbf{-0.03}*** \newline (1.06E-3) &  \textbf{0.01}*** \newline (1.69E-3) & 3.64E-4*** \newline (1.25E-4) & 5.35E-4*** \newline (1.94E-4) & 2.25E-4*** \newline (1.24E-4) \\ \hline
        
        \textbf{Affect/ Negativity Bias} &  \textbf{0.05}*** \newline (6.70E-4) &  \textbf{0.01}*** \newline (3.21E-4) &  \textbf{0.04}*** \newline (6.55E-4) & \textbf{0.02}*** \newline (3.66E-4) &  -4.00E-16*** \newline (2.63E-17) &  -1.51E-18*** \newline (2.20E-18) & \ -2.62E-18*** \newline (3.42E-18) &  -1.39E-18*** \newline (2.19E-18) \\ \hline
        
        \textbf{Cognitive Dissonance} & \textbf{0.09}*** \newline (1.40E-3) & \textbf{0.02}*** \newline (6.71E-4) &  \textbf{0.06}*** \newline (1.37E-3) & \textbf{0.02}*** \newline (7.66E-4) &  0.003*** \newline (1.17E-3) &  -9.88E-5*** \newline 9.81E-5 &  -1.21E-4*** \newline (1.53E-4) & -6.12E-5*** \newline (9.75E-5)  \\ \hline
        
        \textbf{Confirmation Bias} &  \textbf{0.03}*** \newline (1.59E-3) &  0.004*** \newline (7.60E-4) &  \textbf{0.03}*** \newline (1.55E-3) & \textbf{0.01}*** \newline (8.67E-4) &  \textbf{0.05}*** \newline (1.23E-3) & 3.25E-4*** \newline (1.03E-4) &  6.59E-4*** \newline (2.20E-4) &  3.56E-4*** \newline (1.02E-4)  \\ \hline
    \end{tabular}
    \caption{\textbf{Association between Bias Triggers and Tweet Engagement.} Ordinary Least Squares (OLS) correlation coefficients are reported with standard errors in parentheses. Each linear regression model regressed a log-transformed engagement metric against the presence of the eight computational bias triggers. Note: Bolded values indicate practical significance (defined structurally as an effect size magnitude $\geq 0.01$). 
    ***$p<0.01$, **$p<0.05$, *$p<0.1$.}
    \label{tab:regression}
\end{table}

\end{landscape}
\restoregeometry

\end{document}